\documentclass[a4paper,fleqn,usenatbib]{mnras}
\usepackage{amsmath}
\usepackage{amssymb}
\usepackage{cite}
\usepackage{natbib}
\usepackage{graphicx}
\usepackage{xcolor}
\usepackage{txfonts}

\usepackage[T1]{fontenc}
\usepackage{ae,aecompl}

\renewcommand\vec{\bmath}

\usepackage{ulem}

\title[AMICO: optimised detection of galaxy clusters]{%
AMICO: optimised detection of galaxy clusters in photometric surveys
}
\author[Bellagamba, Roncarelli, Maturi, Moscardini]{%
Fabio Bellagamba,$^{1,2}$ Mauro Roncarelli,$^{1,2}$ Matteo Maturi$^3$ and Lauro Moscardini$^{1,2,4}$\\
$^1$Dipartimento di Fisica e Astronomia, Universit\`a di Bologna, via Gobetti 93/2, I-40129 Bologna, Italy\\
$^2$Istituto Nazionale di Astrofisica (INAF) - Osservatorio Astronomico di Bologna, via Gobetti 93/3, I-40129 Bologna, Italy\\
$^3$Zentrum f\"ur Astronomie, Universitat\"at Heidelberg, Philosophenweg 12, D-69120 Heidelberg, Germany\\
$^4$INFN - Sezione di Bologna, viale Berti-Pichat 6/2, I-40127 Bologna, Italy}

\pubyear{2016}

\begin{document}
\label{firstpage}
\pagerange{\pageref{firstpage}--\pageref{lastpage}}
\maketitle

\begin{abstract}
We present AMICO (Adaptive Matched Identifier of Clustered Objects), a new algorithm for the detection of galaxy clusters in photometric surveys. AMICO is based on the Optimal Filtering technique, which allows to maximise the signal-to-noise ratio of the clusters. In this work we focus on the new iterative approach to the extraction of cluster candidates from the map produced by the filter. In particular, we provide a definition of membership probability for the galaxies close to any cluster candidate, which allows us to remove its imprint from the map, allowing the detection of smaller structures. As demonstrated in our tests, this method allows the deblending of close-by and aligned structures in more than $50\%$ of the cases for objects at radial distance equal to $0.5 \times R_{200}$ or redshift distance equal to $2 \times \sigma_z$, being $\sigma_z$ the typical uncertainty of photometric redshifts. Running AMICO on mocks derived from N-body simulations and semi-analytical modelling of the galaxy evolution, we obtain a consistent mass-amplitude relation through the redshift range $0.3 < z < 1$, with a logarithmic slope $\sim 0.55$ and a logarithmic scatter $\sim 0.14$. The fraction of false detections is steeply decreasing with S/N, and negligible at S/N > 5.
\end{abstract}


\section{Introduction}

\footnotetext[1]{\url{http://sci.esa.int/euclid/}}

Clusters of galaxies were historically recognised as overdensities of galaxies in optical images \citep[e.g.][]{1933PNAS...19..591S,1958ApJS....3..211A}. Although large samples of clusters have also been drawn from X-ray observations or through the detection of the Sunyaev-Zel'dovich (SZ) effect on the CMB, photometric observations remain the most promising source of discovery of new clusters of galaxies. This applies especially to the wide ongoing and upcoming photometric surveys such as the Kilo Degree Survey \citep[KiDS,][]{2013Msngr.154...44J}, the Dark energy Survey \citep[DES,][]{2005astro.ph.10346T}, Pan-STARRS \citep{2002SPIE.4836..154K}, the Large Synoptic Survey Telescope \citep[LSST,][]{2009arXiv0912.0201L} and the European Space Agency Cosmic Vision mission \textit{Euclid}\footnotemark\ \citep{2011arXiv1110.3193L}. On the other hand, the detection of clusters through photometric data requires non-obvious procedures to distinguish genuine physical groups from clumps generated by chance alignments and non-collapsed overdensities due to the matter distribution on larger scales. 

Because of these challenges, and following the evolution in the quality and the amount of data to be analysed, a multitude of automated methods have been developed in the last decades. A possible (but not rigid) classification can be made according to the kind of information the algorithm uses to perform the detection. A first class of methods are purely geometrical, i.e. they try to group galaxies from their positions (including photometric redshifts, when available). These methods do not consider explicitly the galaxy observational properties (i.e. magnitudes and colors) in the detection process, although they are accounted for in the photo-z measurement. The first automated approaches, such as those by \citet{1985ApJS...57...77S} and \citet{1992MNRAS.258....1L}, performed a Gaussian smoothing of galaxy counts in cells. More recent and sophisticated attempts include Friends of Friends algorithms \citep[e.g.][]{2004MNRAS.349..425B,2007A&A...463..853T,2012ApJS..199...34W} and those based on Voronoi tessellation \citep[e.g.][]{1999ASPC..176..108R,2011ApJ...727...45S}. 

Another group of methods is instead based on known observational properties of the galaxy population in clusters, and in particular on the presence of a dominant passive component. In practice, clusters are searched as overdensities of galaxies with a similar (red) colour. The evolution of the observed colour for this galaxy population encodes the redshift information and makes photo-zs redundant in this kind of analysis. This way of searching for clusters was named Cluster Red Sequence method by \citet{2000AJ....120.2148G}, and then evolved in the MaxBCG algorithm \citep{2005ApJ...633..122H,2007ApJ...660..221K} and finally in redMaPPer \citep{2014ApJ...785..104R,2016ApJS..224....1R} and RedGOLD \citep{2016MNRAS.455.3020L}. 

A different approach is the one of Matched Filters (MF), which dates back to \citet{1996AJ....111..615P}. This approach requires an a-priori definition of the cluster model, usually made up of a radial profile and a luminosity function, which may vary also with redshift: clusters are searched as patterns in the data that resemble the model and are not likely to be produced by random fluctuations. This method evolved in different directions: the Adaptive Matched Filter \citep{1999ApJ...517...78K, 2002ApJ...574...24W, 2008ApJ...676..868D}, 3D-MF \citep{2010MNRAS.406..673M}, and the Bayesian Cluster Finder \citep{2012MNRAS.420.1167A} which, differently from the others, includes galaxy colours and the presence of a BCG in the likelihood evaluation.

Optimal Filtering is a general technique to extract and measure an expected signal in a data-set affected by a noisy background. In astronomy, it was originally applied to weak lensing detection of galaxy clusters by \citet{2005A&A...442..851M} and \citet{2007A&A...471..731P}, and to X-ray and SZ detections in \citet{2008A&A...483..389P}. In \citet{2011MNRAS.413.1145B}, it was shown that in case of photometric detections, if one assumes that the background is homogeneous, the Optimal Filter actually corresponds to a Matched Filter. A more recent application to photometric data of the same method can be found in \citet{2017A&A...598A.107R}. Differently from other methods, the formalism of Optimal Filtering is very generic and allows to consider (or neglect) in the model any galaxy property, such as magnitudes or colours, the presence of a BCG, or even photometric and morphological classifications. It is the algorithm itself which selects the most relevant properties for the cluster detection in the available data-set and thus gives appropriate weights to the galaxies in the catalogue. The extreme flexibility of the Optimal Filtering method makes it suitable for surveys which span a wide range in redshift, differently from other methods which are tailored to a specific kind of data. Moreover, this formalism allows to combine in a consistent way optical data with other observables such as weak gravitational lensing, SZ and X-rays observations.

In this paper we present AMICO (Adaptive Matched Identifier of Clustered Objects), an improved Optimal Filtering algorithm for cluster detection. Apart from many improvements over the version described in \citet{2011MNRAS.413.1145B} in terms of usability, robustness and efficiency, it presents new features in the detection procedure itself. In particular, in this work we will focus on a new method to translate the map derived from the application of the filter to a list of cluster detections, which is the ultimate and the most important output of such an algorithm. The basic idea is to iteratively remove the imprint of candidate clusters from the map, allowing for the detection of other objects at lower signal-to-noise ratio in the surroundings. This becomes a critical point when dealing with surveys which are deep enough to detect tens of clusters/groups per square degree, making cluster detection severely limited by confusion. A crucial quantity in this procedure is the membership probability that is calculated for all possible cluster members after each detection. In practice, each galaxy is initially considered part of the background distribution and, as the detection procedure goes on, may be gradually attributed probabilistically to one or more detections. A deblending technique allowing the removal of the candidate clusters has already been applied to SZ detections, but only through the cluster template, instead of using the likely members of the cluster \citep{2008A&A...483..389P}. This new approach leads to more accurate and self-consistent results. An other important feature of AMICO is the local estimation of the background population which, as we will see, is a key ingredient in getting an unbiased mass proxy for the detections. 

A critical point in all filtering methods is the definition of the model or template, that is the expected appearance of the clusters in the given data-set. On one side, using the knowledge about clusters helps the detection increasing the signal-to-noise ratio of poor or distant structures. On the other side, when more and more information is included, it may induce an ``over-specialization'' of the filter which is going to target a very specific type of objects leading to a non-straightforward selection function for the cluster sample, or it may produce sub-optimal results if the filter is applied in regions of wavelength or redshift where knowledge about clusters is limited. As stated above, AMICO employs a very general structure, and the cluster model is considered an input parameter of the algorithm. AMICO has the potential to iteratively perform the extraction and the refinement of the model for the cluster from the available data starting from simple assumptions, but we leave this topic for a future paper. In the following, we will assume to know the model that describes the expected galaxy distribution in clusters at each redshift. 

Many of the key ingredients in the definition of the detection performance of AMICO and of other cluster finders depend, of course, on survey specific quantities, such as its depth, the magnitude errors, the quality of photometric redshifts and the accuracy of their error modeling. Since in this paper we do not aim to make any forecast for specific applications of AMICO to current or future data-sets, we instead focus in providing a general presentation ought to be used as a benchmark for results on real data. For this reason, and to keep the analysis as simple as possible, in the catalogues used in this paper the photometric redshifts are modelled as simple Gaussian distributions with a realistic width. Nevertheless, as described in the following Section, AMICO can deal with arbitrary probability redshift distributions as provided by photometric redshift algorithms.

The paper is organised as follows. In Section \ref{sect:filter} we give an overview of the basic principles of Optimal Filtering and we describe the relevant quantities for its application to photometric data. In Section \ref{sect:selecting} we outline the newly implemented procedure to iteratively extract clusters and their members from the map produced by the application of the filter to the data. In Sections \ref{sect:ideal} we perform different tests on fields with clusters in specific configurations to highlight the capability of the algorithm in disentangling clusters which are close-by or aligned along the line of sight. In Section \ref{sect:real_mocks} we perform the cluster detection in a more realistic field, where clusters have a more diverse population and are embedded in a realistic, non homogeneous, background. Finally, in Section \ref{sect:concl} we summarise the results of our work and make some concluding remarks.

\section{Optimal Filtering of the data} \label{sect:filter}
Optimal Filtering is a general technique to extract and measure an expected signal in a data-set which is affected by a noisy background. We refer to \citet{2005A&A...442..851M} and \citet{2011MNRAS.413.1145B} for a more complete introduction about Optimal Filters and their application to cluster detection in photometric surveys, respectively. In this Section, we review the relevant  quantities for the discussion of the specific implementation of the algorithm in AMICO.

\subsection{General principles}\label{sect:filter_constr}
The basic assumption of Optimal Filtering is that we can describe a set of data $D$ as the sum of a signal component described by a model $M$ with an unknown normalisation $A$ and a noise component $N$. In the case of a photometric catalog, the data are represented by the galaxy density $D(\vec \theta, \vec m, z)$ as a function of angular position in the sky $\vec \theta$, of magnitudes $\vec m$ and of redshift $z$. We can write it as 
\begin{equation}\label{eq:data}
D(\vec \theta, \vec m, z) = A(\vec \theta_c, z_c) \times M_c(\vec \theta - \vec \theta_c, \vec m, z) + N (\vec m, z) .
\end{equation}
The model $M_c(\vec \theta - \vec \theta_c, \vec m, z)$ is the expected galaxy distribution of a cluster centred in ($\vec \theta_c, z_c$), the amplitude $A(\vec \theta_c, z_c)$ is the normalisation of the cluster galaxy distribution, and the noise $N(\vec m, z)$ is the field galaxy distribution. In this formalism, the array $\vec m$ may include not only photometric magnitudes, but also any other observable available for the galaxies in the catalog, such as morphological type or ellipticity. 
Here we will assume the distribution of the field galaxies to be completely random with a uniform mean, i.e. the power spectrum of the large-scale structure is neglected. An improvement over this simplistic assumption will be described in Section \ref{sect:localback}. 

Following the general theory of Optimal Filtering, the optimal linear estimate of the amplitude $A$ of the signal component in the position $(\vec \theta_c, z_c)$ is obtained by filtering the data $D$ with a suitable function $\Psi_c$
\begin{equation}\label{eq:generic_amplitude}
A(\vec \theta_c, z_c) = \alpha^{-1}(z_c) \int \Psi_c(\vec \theta - \vec \theta_c, \vec m, z) D(\vec \theta, \vec m, z)\ d^2 \theta\ d^n m\ dz - B(z_c) ,
\end{equation}
where $\alpha$ is a normalisation constant given by
\begin{equation}\label{eq:generic_alpha}
\alpha(z_c) = \int \Psi_c^2(\vec \theta - \vec \theta_c, \vec m, z) {N(\vec m, z_c)}\ d^2 \theta\ d^n m\ dz ,
\end{equation}
where $n$ is the number of dimensions of the space describing the cluster properties $\vec{m}$, and $B$ is the background component to be subtracted. Under the assumption that the noise is uniform and is produced by random Poissonian counts of galaxies, the Optimal Filter is given by the ratio of the model $M_c$ over the data noise $N$:
\begin{equation}
\Psi_c(\vec \theta - \vec \theta_c, \vec m, z) = M_c(\vec \theta - \vec \theta_c, \vec m, z) / N (\vec m, z_c).
\end{equation}
It is possibile to demonstrate that the estimate of Equation \ref{eq:generic_amplitude} is optimal in the sense that it is unbiased and with the minimum possible variance.

\subsection{Applying the filter}\label{sect:filter_phot}
Dealing with a catalogue of galaxies, it is convenient to rewrite Eq. \ref{eq:generic_amplitude} as
\begin{equation}\label{eq:real_amplitude}
A(\vec \theta_c, z_c)  = \alpha^{-1}(z_c)\ S(\vec \theta_c, z_c)  - B(z_c) ,
\end{equation}
where $S$, defined as
\begin{equation}\label{eq:rough_amplitude}
S(\vec \theta_c, z_c) = \sum_{i=1}^{N_{gal}} \frac {M_c(\vec \theta_i - \vec \theta_c, \vec m_i)\ p_i(z_c)}{N(\vec m_i,z_c)} ,
\end{equation}
is the analogue of the integral of Eq. \ref{eq:generic_amplitude} for discrete data, and the subscript $i$ runs over the galaxy catalogue. In the model $M_c$ of a cluster at redshift $z_c$, we are now omitting the redshift distribution of the members, in favour of the photometric redshift distribution $p_i(z)$ for each galaxy. In fact, the intrinsic dispersion in redshift of the cluster members is negligible in comparison with the measurement uncertainty.  The units of the model $M_c(r, \vec m)$ are then $\mathrm{mag^{-n}} \deg^{-2}$, the noise $N(\vec m,z)$ instead has units $\mathrm{mag^{-n}} \deg^{-2} \mathrm{z^{-1}}$, as the data $D(\vec \theta, \vec m, z)$.

The normalization $\alpha$ ensures that the amplitude $A$ is the measure of the cluster signal in units of the model $M_c$. Eq. \ref{eq:generic_alpha} then becomes
\begin{equation}\label{eq:alpha}
\alpha(z_c) = \int \frac {M_c^2(\vec \theta - \vec \theta_c, \vec m)\ q^2(z_c, z)} {N(\vec m,z_c)}\ d^2\theta\ d^n m\ dz.
\end{equation}
In the previous equation, we introduced $q(z_c, z)$, which is the typical redshift probability distribution for a galaxy which lies at redshift $z_c$. If no a-priori information is available, this distribution can be computed from the data as 
\begin{equation}\label{eq:mean_pz}
q(z_c,z) = \left(\sum_{i=1}^{N_\text{gal}} p_i(z_c)  \right)^{-1} \sum_{i=1}^{N_\text{gal}} p_i(z-z_c+z_\text{peak,i})\ p_i(z_c) ,
\end{equation}
where $z_\text{peak,i}$ is the most probable redshift for the $i$-th galaxy. Equation \ref{eq:mean_pz} is a sort of weighted mean $p(z)$ for galaxies whose true redshift is $z_c$ computed in this way: each galaxy probability distribution is weighted with its value at redshift $z_c$ and then is shifted so that it peaks at $z_c$. Instead of a generic $q(z_c,z)$ for all objects, it is possible to compute Equation \ref{eq:mean_pz} for different classes of objects (e.g. as a function of magnitudes) and take this into account in the integral of Equation \ref{eq:alpha} and in the following.

The same statistics on the redshift distribution can be used to calculate an analytical estimate of the background component to be subtracted, which is
\begin{equation} 
B(z_c) = \alpha^{-1}(z_c) \beta(z_c) ,
\end{equation}
where $\beta(z_c)$ is the expectation value of $S(\vec \theta_c, z_c)$ if the galaxy distribution corresponds to the field component only:
\begin{eqnarray}\label{eq:beta}
\beta(z_c) &=& \int \frac {M_c(\vec \theta - \vec \theta_c, \vec m)\ q(z_c, z)} {N(\vec m,z_c)} {N(\vec m,z_c)}\ d^2\theta\ d^n m\ dz = \notag \\
&=& \int M_c(\vec \theta - \vec \theta_c, \vec m)\ q(z_c, z)\ d^2\theta\ d^n m\ dz .
\end{eqnarray}
The subtraction of $B(z_c)$ in Equation \ref{eq:real_amplitude} ensures that $A$ = 0 when the galaxy distribution around a given position corresponds to the field one, namely $D(\theta_c,z_c)=N(\theta_c,z_c)$. By construction, $\beta(z_c)$ is also the total number of galaxies in the cluster model at redshift $z_c$.  

It is important to stress that the treatment of photometric redshifts described here is completely generic, and does not assume any specific shape or accuracy of the p(z) distribution. This is different with respect to other Matched Filter algorithms, such as \citet{2008ApJ...676..868D} and \citet{2010MNRAS.406..673M}. Even in the absence of any redshift information one can apply the formalism described here by simply assuming a flat probability distribution in the considered redshift range.

\subsection{Variance of the amplitude estimate}
Following the assumptions described in the previous Sections, one can derive an analytical estimate of the uncertainty of $A(\vec \theta_c, z_c)$. This is given by
\begin{equation}\label{eq:uncertainty}
\sigma^2_A (\vec \theta_c, z_c)= \alpha^{-1}(z_c) + A(\vec \theta_c, z_c)\ \frac {\gamma(z_c)} {\alpha^2(z_c)} ,
\end{equation}
where the first component derives from the background fluctuations, the second one accounts for the shot-noise contribution of the actual structures present in the data as measured with the filter. Here, the quantity $\gamma$ is given by
\begin{equation}\label{eq:gamma}
\gamma(z_c) = \int \frac {M_c^3(\vec \theta - \vec \theta_c, \vec m)\ q^3(z_c, z)} {N^2(\vec m,z_c)}\ d^2\theta\ d^n m\ dz .
\end{equation}
By construction, Equation \ref{eq:uncertainty} is meaningless when $A$ is negative, but this is not important as regions with negative $A$ have a lower-than-average galaxy density, and so are negligible in the cluster detection.

\subsection{Likelihood}
Given the description of the data $D$ as the sum of a signal component $M$ with normalisation $A$ and a noise component $N$ as in Equation \ref{eq:data}, their likelihood is
\begin{equation}\label{eq:generic_like}
\mathcal{L}(\vec \theta_c, z_c) = - \int \frac {\{D - [A M_c(\vec \theta - \vec \theta_c, \vec m)+ N(\vec m,z_c)]\}^2} {N(\vec m,z_c)} d^2\theta\ d^n m\ dz  .
\end{equation} 
Combining the previous Equation with the amplitude estimate given by Equation \ref{eq:real_amplitude}, one can show that 
\begin{equation}\label{eq:likelihood}
\mathcal{L}(\vec \theta_c, z_c) = \mathcal{L}_0 +  A^2(\vec \theta_c, z_c)\ \alpha(z_c) ,
\end{equation}
where $\mathcal{L}_0$ is a constant that does not depend on the position. Differently from AMF \citep{1999ApJ...517...78K,2008ApJ...676..868D}, we do not consider a Poissonian likelihood, but use instead a Gaussian likelihood, coherently with the construction of the Optimal Filter.

At a given redshift, maxima of $A$ are also maxima of the likelihood. Intuitively, this depends on the fact that, the bigger is the cluster, the bigger is the improvement in the likelihood when the cluster model is centred in that position. However, as noted in \citet{2011MNRAS.413.1145B}, the likelihood alone does not distinguish between overdensities ($A > 0$), which we are interested in, and underdensities with negative $A$. In the following, to perform the detection we will combine this information with the signal-to-noise ratio $A/\sigma_A$, which is meaningful only when $A$ is positive.

To summarise, following Eqs. \ref{eq:real_amplitude}, \ref{eq:uncertainty} and \ref{eq:likelihood}, we can calculate an estimate of the amplitude of the candidate cluster, its uncertainty and the likelihood on a 3D grid of positions in the sky and redshift. We can then derive cluster detections from this map by selecting ($\vec \theta_c, z_c$) locations with large signal-to-noise values and large likelihood. We will detail our procedure to extract detections from the map in Section \ref{sect:selecting}.

We note that, differently from other works \citep{2008ApJ...676..868D,2012MNRAS.420.1167A} that center their cluster candidates on galaxy positions, we use a regular grid in angles and redshift. This choice is helpful for the optimisation of the extraction of cluster candidates presented in Section \ref{sect:selecting}. If needed, it is always possible to select the most likely central galaxy a posteriori.

\subsection{Masked areas}
It is common in galaxy surveys to have field regions with missing data, mainly due to foreground sources (e.g. stars). This is usually handled with the definition of suitable masks. In our formalism, we can account for masked regions by computing the constants $\alpha$, $\beta$ and $\gamma$ as a function of $\vec x_c$ (and not just $z_c$), performing the integrals in Eqs. \ref{eq:alpha}, \ref{eq:beta} and \ref{eq:gamma} only on the available (unmasked) area around the center. The derived values of $A$ are then comparable, because they are computed assuming the same model and the noise distribution, but considering the different available area for the amplitude estimate. By construction, if the galaxy distribution is the same, the signal-to-noise ratio of an amplitude estimate derived from a partially masked area will be lower than the one obtained when the full area is completely available to observations.

\subsection{Local background}\label{sect:localback}
In the previous sections the background component $N(\vec m, z_c)$ has been considered to have a constant mean over the surveyed area. With this approximation, for a given redshift $z_c$, the contribute of the field population to the amplitude is random with mean equal to $B(z_c)$ and is subtracted as in Equation \ref{eq:real_amplitude}. In reality, the large-scale structure of the Universe will produce correlations in the density of galaxies on scales which are larger than the typical cluster scale. In order to appropriately remove the background contribution to the amplitude estimate $A$, AMICO applies a local correction $f(\vec \theta, z_c)$ to $N(\vec m, z_c)$. For every redshift, this is computed by producing a map of the density of galaxies as a function of the angular coordinates. A k-sigma clipping is applied to this map to remove obvious peaks (likely to be produced by clusters and not by the field distribution) and then a smoothing is performed over a scale which is significantly larger than the one of the filter. Finally, by dividing this map by its mean value, we obtain a correction term $f(\vec \theta, z_c)$ for each position. 

It is easy to show that one can account for the local background correction a-posteriori. In fact, after applying the filter to the galaxy population using the mean $N(m,z_c)$ (Equation \ref{eq:rough_amplitude}), the relevant quantities for the corrected background density $f(\vec \theta, z_c)N(m,z_c)$ can can be calculated in the following way:
\begin{equation}\label{eq:localback_amplitude}
A^\prime(\vec \theta_c, z_c) = \frac {S(\vec \theta_c, z_c)  - f(\vec \theta, z_c) \beta(z_c)}{\alpha(z_c)} \;\; ,
\end{equation}
\begin{equation}\label{eq:localback_uncertainty}
\sigma^2_{A^\prime}(\vec \theta_c, z_c) =  \frac {f(\vec \theta, z_c)} {\alpha(z_c)} + A^\prime(\vec \theta_c, z_c)\ \frac {\gamma(z_c)} {\alpha^2(z_c)} \;\; ,
\end{equation}
\begin{equation}\label{eq:localback_likelihood}
\mathcal{L^\prime}(\vec \theta_c, z_c) = \mathcal{L}_0 +  \frac {A^2(\vec \theta_c, z_c)\ \alpha(z_c)}{f(\vec \theta, z_c)}\;\; .
\end{equation}
In this step there are no approximations, the very same results would be obtained by accounting for the local correction from the beginning. We opted for this procedure for efficiency reasons.

\subsection{Model}
In the formalism described above, the model is an input parameter, that describes the expected galaxy distribution in clusters as a function of redshift. In principle, one can construct it in several ways, for example by binning members of known clusters in magnitude and radial distance. For numerical stability and continuity, it is often preferred to use analytic expressions to describe the model. As a cluster galaxy distribution can be considered the sum of different components (e.g. central galaxy and satellites, red and blue populations), the model can be written in a generic form as
\begin{equation}
M_c(r, \vec m) = \sum_i \Psi_i(r) \Phi_i(\vec m)\;\; ,
\end{equation}
where the index $i$ runs over the considered galaxy populations and $\Psi_i(r)$ and $\Phi_i(\vec m)$ are the radial and magnitude distributions for the $i$-th component, respectively.

As a representative example of a practical choice for the model, we describe here the model used for the tests presented in Sections \ref{sect:ideal} and \ref{sect:real_mocks} of this paper. In this case, we consider separately the distribution of central and satellite galaxies, and we take into account only one observing band. The central galaxies are modelled with a Gaussian distribution in magnitude. In the radial distribution, they are described by a step distribution larger than zero only in the most inner bin, as they lie at the center of the halos by construction. For the satellites' magnitude distribution, we use a Schechter function \citep{1976ApJ...203..297S}:
\begin{equation}\label{eq:schechter}
\Phi(m) = C \times 10^{-0.4(m-m_\star)(\alpha+1)} \exp{(-10^{-0.4(m-m_\star)})},
\end{equation}
where $m_\star$ is the typical magnitude, $\alpha$ controls the faint-end slope and $C$ is the normalisation. For the satellites' radial profile we use a NFW distribution \citep{1997ApJ...490..493N} 
\begin{equation}\label{eq:nfw}
\Psi(r) = \frac {B}{\displaystyle \frac {r}{r_s} \biggl(1+\frac{r}{r_s}\biggr)^2},
\end{equation}
where the scale radius is $r_s = R_{200}/c$, $c$ is the concentration parameter and $B$ is the normalisation. The value of the parameters as a function of redshift is derived from mock catalogues as described in Section \ref{sect:model}.

\section{Selecting the cluster candidates}\label{sect:selecting}
The principles of Optimal Filtering, applied to the photometric galaxy survey data, lead to the construction of a 3D map of amplitude estimates $A(\vec \theta, z_c)$, as defined by Equation \ref{eq:real_amplitude}. Each value of this map represents an estimate of the amplitude of a cluster signal, if a cluster is actually centred in that position. It is non-trivial to move from the amplitude map to the list of likely clusters, which is the desired output of a detection algorithm. In the following, we will outline how AMICO proceeds in this task. In particular, we will describe the {\it cleaning} procedure, which aims at removing the signal of the most significant objects from the map in the attempt to detect smaller objects which might be blended with them. This is one of the most important novelties introduced in AMICO with respect to traditional matched-filter algorithms. We will see in Sections \ref{sect:deblending} and \ref{sect:cleaning_z} a practical demonstration of its power.

After the 3D map of amplitude is built as described in Section \ref{sect:filter_phot}, the algorithm proceeds iteratively. The total number of detections $N_{det}$ depends on $(S/N)_{min}$, the minimum signal-to-noise ratio one wants to achieve. For $j$ that runs from 1 to $N_{det}$ the following steps are made.
\begin{enumerate}
\item All the pixels with a S/N below the defined threshold $(S/N)_{min}$ are discarded.
\item Among the remaining pixels, the one with the highest likelihood is selected. This identifies the $j$-th cluster detection.
\item Some relevant properties of the detection are saved, such as center position ($\vec \theta_j, z_j$), amplitude $A_j \equiv A(\theta_j, z_j)$, signal-to-noise ratio $A_j/ \sigma_{A_j}$ and a first estimate of the number of members $N_j$, given by $N_j = A_j \times \beta(z_j)$.
\item For each galaxy whose sky coordinates lies close to $\theta_j$, a membership probability $P(i \in j)$ is computed as detailed in Section \ref{sect:members}.
\item The imprint of its potential members is removed from the map, as it will be detailed in Section \ref{sect:cleaning}. 

\end{enumerate}
Once step (v) is completed, all the values of amplitude, likelihood and S/N are updated and do not contain anymore the contribution of the $j$-th detection (and of the previous ones). The process is repeated iteratively until after step (i) no pixel is left.

\subsection{Membership assignment}\label{sect:members}
We now describe the details regarding the step (iv) of our procedure. Given the $j$-th detection at position $\vec \theta_j$ and redshift $z_j$ with amplitude $A_j$, the probability $P( i\in j )$  of the $i$-th galaxy of being a member of the $j$-th detected object is defined as
\begin{equation}\label{eq:member_prob}
P( i\in j ) \equiv \frac {A_j M_j(\vec \theta_i - \vec \theta_j, \vec m_i) p_i(z_j)}{A_j M_j(\vec \theta_i - \vec \theta_j, \vec m_i) p_i(z_j) + N(\vec m_i,z_j)} ,
\end{equation}
i.e. the ratio between the galaxy density associated to the cluster component and the total one.
This definition is strictly correct only if we consider one cluster at a time, while in reality several galaxies will have a positive probability to belong to more than one structure. In principle, Eq. \ref{eq:member_prob} should be modified to
\begin{equation}\label{eq:member_prob_correct}
P( i\in j ) = \frac {A_j M_j(\vec \theta_i - \vec \theta_j, \vec m_i) p_i(z_j)}{\sum_{k=1}^{N_{det}} A_k M_k(\vec \theta_i - \vec \theta_k, \vec m_i) p_i(z_k) + N(\vec m_i,z_j)} ,
\end{equation}
where the sum at the denominator runs over all the detected structures. However, even if formally correct, the application of this equation in the process outlined above is unfeasible. In fact, the probability must be computed when the $j$-th detection is found and no information is available on the presence of other nearby clusters candidates which will be subsequently detected. Thus, we consider one cluster at a time in the total density, an assumption which is coherent with the measurement of $A$ made via Eq. \ref{eq:real_amplitude}. 

We take into account previous memberships assigned to the same galaxy by introducing the field probability $P_{f,i}$, which is defined as
\begin{equation}\label{eq:field_prob}
P_{f,i} = 1 - \sum_{k=1}^{j-1} P( i\in k ) .
\end{equation}
The field probability for a galaxy is 1 at the beginning of the process, and decreases while the galaxy is progressively attributed to new detected objects. Intuitively, it is a measure of the fraction of the galaxy which is still available for further associations. Then, Eq. \ref{eq:member_prob} becomes
\begin{equation}\label{eq:member_prob_corr}
P( i\in j ) = P_{f,i}  \times \frac {A_j M_j(\vec \theta_i - \vec \theta_j, \vec m_i) p_i(z_j)}{A_j M_c(\vec \theta_i - \vec \theta_j, \vec m_i) p_i(z_j) + N(\vec m_i,z_j)} .
\end{equation}
This modification also ensures that the sum of the membership probabilities of each galaxy is always $\leq 1$.

\subsection{Cleaning the detections}\label{sect:cleaning}

Once the probability of a galaxy to belong to a given detected object is calculated, in step (v) of the process summarised above we can take this knowledge into account when determining the following detections. We do so by removing from all the pixels the contribution of possible cluster members weighted by $P( i\in j )$. This effectively removes the imprint of the detected cluster from all the pixels.
In practice, Equation \ref{eq:rough_amplitude} becomes
\begin{equation}\label{amplitude_corr}
S(\vec \theta_j, z_j) = \sum_{i=1}^{N_{gal}} P_{f,i} \frac {M_j(\vec \theta_i - \vec \theta_j, \vec m_i) p_i(z_j)}{N(\vec m_i,z_j)} .
\end{equation}

From the computational point of view, the cleaning process forces the code to run through all the galaxies every time a new detection is found, and modify the value of the map of $S(\theta_j,z_j)$ in its surroundings. Moreover, following eqs. \ref{eq:uncertainty} and \ref{eq:likelihood} also the values of $\sigma_A$ and $\mathcal{L}$ must be updated accordingly. This makes the cleaning the computational \textit{bottleneck} of AMICO.

\begin{figure}
\centering
\begin{tabular}{c}
 \includegraphics[width=.85\columnwidth]{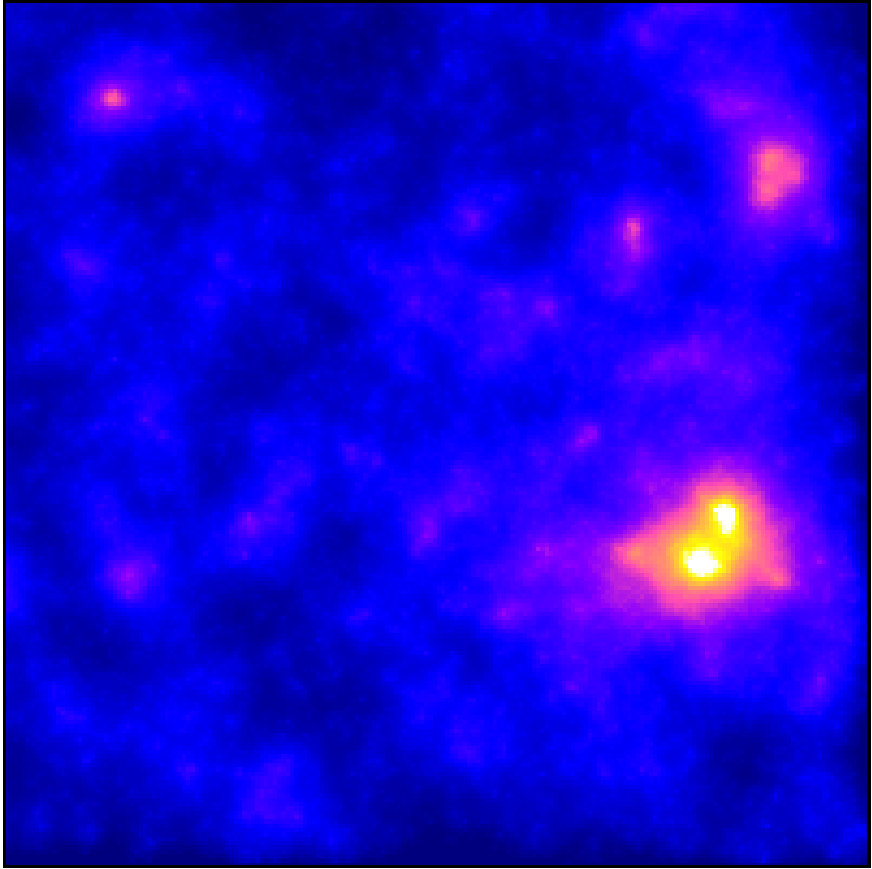}\\
 \includegraphics[width=.85\columnwidth]{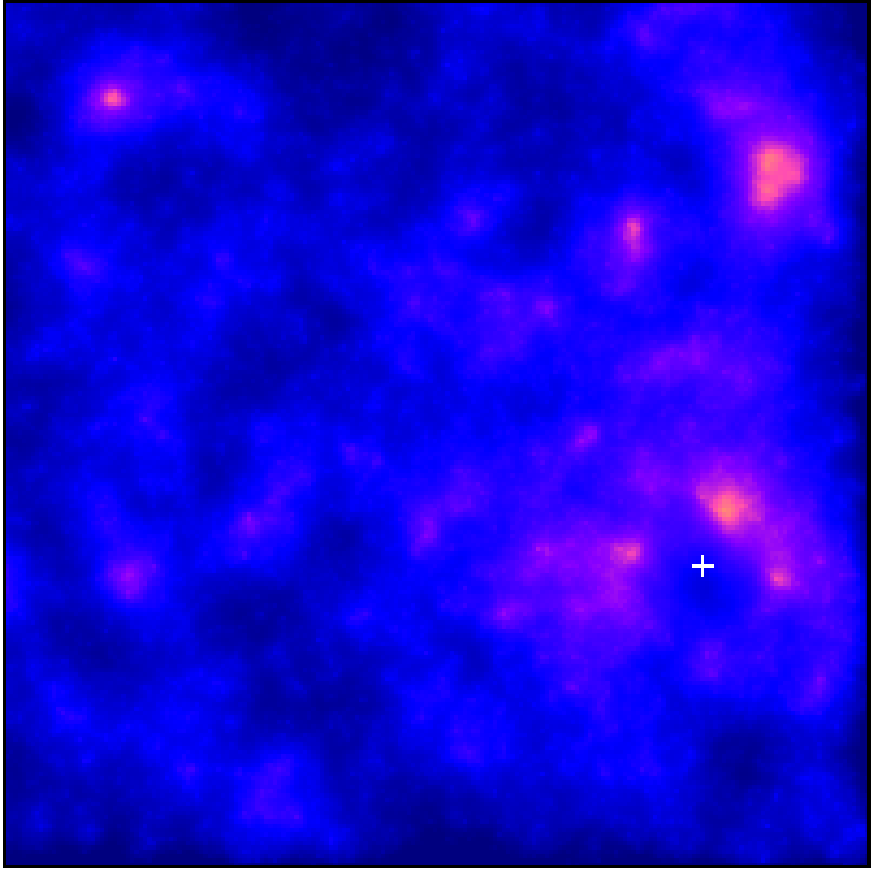}\\
 \includegraphics[width=.85\columnwidth]{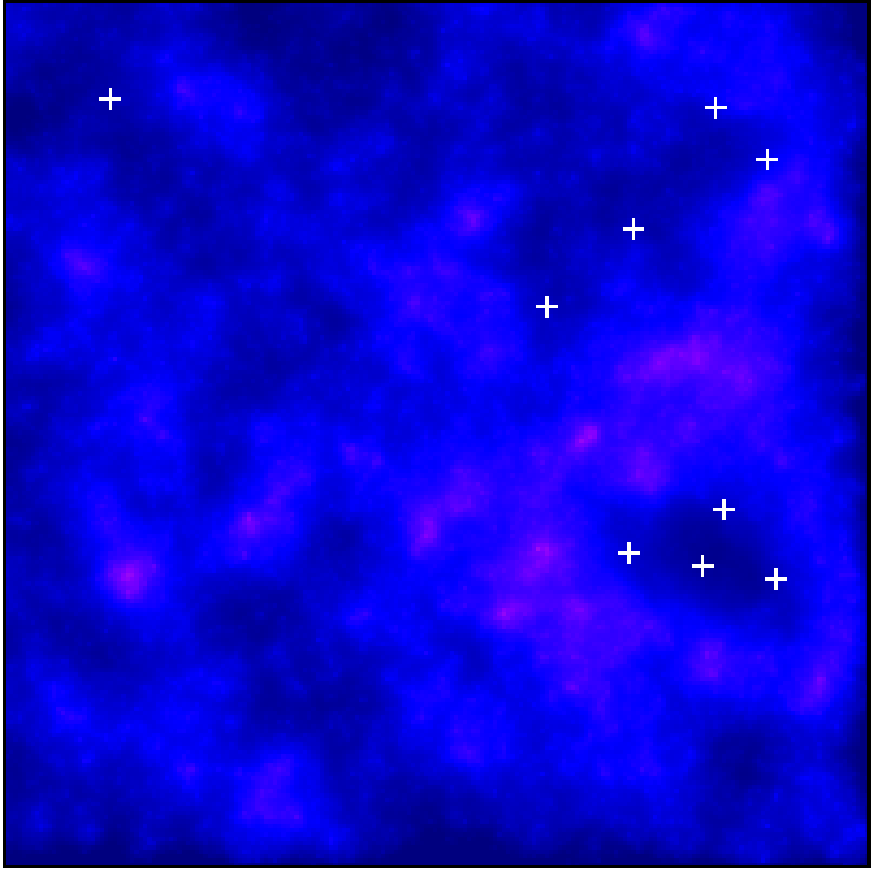}\\
\end{tabular}
 \caption{Amplitude map of $1 \times 1 \deg^2$ at redshift $z$ = 0.33 from one of the fields analysed in Section \ref{sect:real_mocks}. Top panel: initial map produced by the algorithm. Central panel: map after the pixel with the highest S/N has been marked as a detection and its cleaning performed. Bottom panel: map after all the detections with S/N > 7 have been selected and cleaned. White crosses indicate the position of cluster detections removed from the map.}
 \label{fig:cleaning_map}
\end{figure}

An example of how the cleaning procedure modifies the initial amplitude map is shown in Figure \ref{fig:cleaning_map}, from the analysis on the mock galaxy catalogues presented in Section \ref{sect:real_mocks}. 

\section{Tests on ideal mocks}\label{sect:ideal}
In order to verify the performance of the algorithm outlined in the previous sections, we perform two sets of tests, which are distinguished by the kind of mock catalogues used. The first set of tests is run over so-called \textit{ideal} mocks, where we place clusters in known positions over a background population with constant mean density. These tests are useful to show how the algorithm works in a completely controlled environment, where all the assumptions of the Optimal Filtering described in Section \ref{sect:filter} are perfectly met. The second set of tests is run on \textit{cosmological} mocks, where the galaxy distribution, derived from N-body numerical simulations and semi-analytic modelling, is more similar to the one we see in real observations. These tests show the algorithm at work in a more complex environment, where galaxy clusters and groups may have different shapes, luminosity functions, are embedded in the correlated large-scale structure with non-negligible chances of random alignments. All of this makes the identification of clusters more difficult, and the estimation of their properties more uncertain. This second set of tests will be described in Section \ref{sect:real_mocks}. 

\subsection{Simulation}\label{sect:simul}
The starting point for our tests are the public mock catalogues described in \citet{2008A&A...478..299M}. We use 50 catalogues, $2 \times 2 \deg^2$ each, which were originally meant to mimic the VVDS survey \citep{2003SPIE.4834..173L}. They were obtained by applying the prescriptions of \citet{2007MNRAS.375....2D} to the dark-matter halo merging trees extracted from the Millennium simulation \citep{2005Natur.435..629S}. The catalogues are complete in apparent magnitude up to $I$ = 25 for redshift $z$ > 0.3. In this Section, we extract a cluster model and a noise distribution from these mocks and we use them to create new \textit{ideal} mocks. The analysis of the mocks in their original form is discussed in the next Section. The cosmology adopted in the simulation of the mocks is a flat $\Lambda \mathrm{CDM}$ model, with $\Omega_\text{m}$ = 0.25, $\Omega_\Lambda$ = 0.75, $\sigma_8$ = 0.9 and $\text{H}_0$ = 0.73 km $\text{s}^{-1}$ $\text{Mpc}^{-1}$.

\subsection{Model and background}\label{sect:model}

The catalogues provide us a halo identification made on the outputs of the simulation. Galaxies which lie at the center of haloes are marked and for each halo its mass $M_{200}$\footnotemark\ is provided. This allows us to build the model distribution $M_c(r,m)$ which describes the typical galaxy content of a cluster of mass $\sim 10^{14} M_{\odot}/h$ at redshift $z_c$. To this end, we first select as cluster members the galaxies which are at distance $< 1.35 R_{200} \sim R_{100}$ from the central galaxy of a cluster with $13.8 < \log M_{200}/(M_{\odot}/h) < 14.2$. Then, we build the radial distribution (in units of $R_{200}$) and the magnitude distribution for the cluster satellites in bins of redshift, weighing each galaxy by $w_i =   [M_{200} / (10^{14} M_{\odot}/h)]^{-1}$. This weight is used to keep the desired normalisation for the cluster model. Finally, in each redshift bin we fit the derived average radial distribution with a NFW function (Eq. \ref{eq:nfw}) and the average magnitude distribution with a Schechter function (Eq. \ref{eq:schechter}).

\footnotetext[2]{The mass of the haloes is provided in terms of $M_{200}$, which is the mass contained in a sphere with a density 200 times larger than the critical one. The radius of this sphere is $R_{200}$.}

\begin{figure}
 \includegraphics[width=\columnwidth]{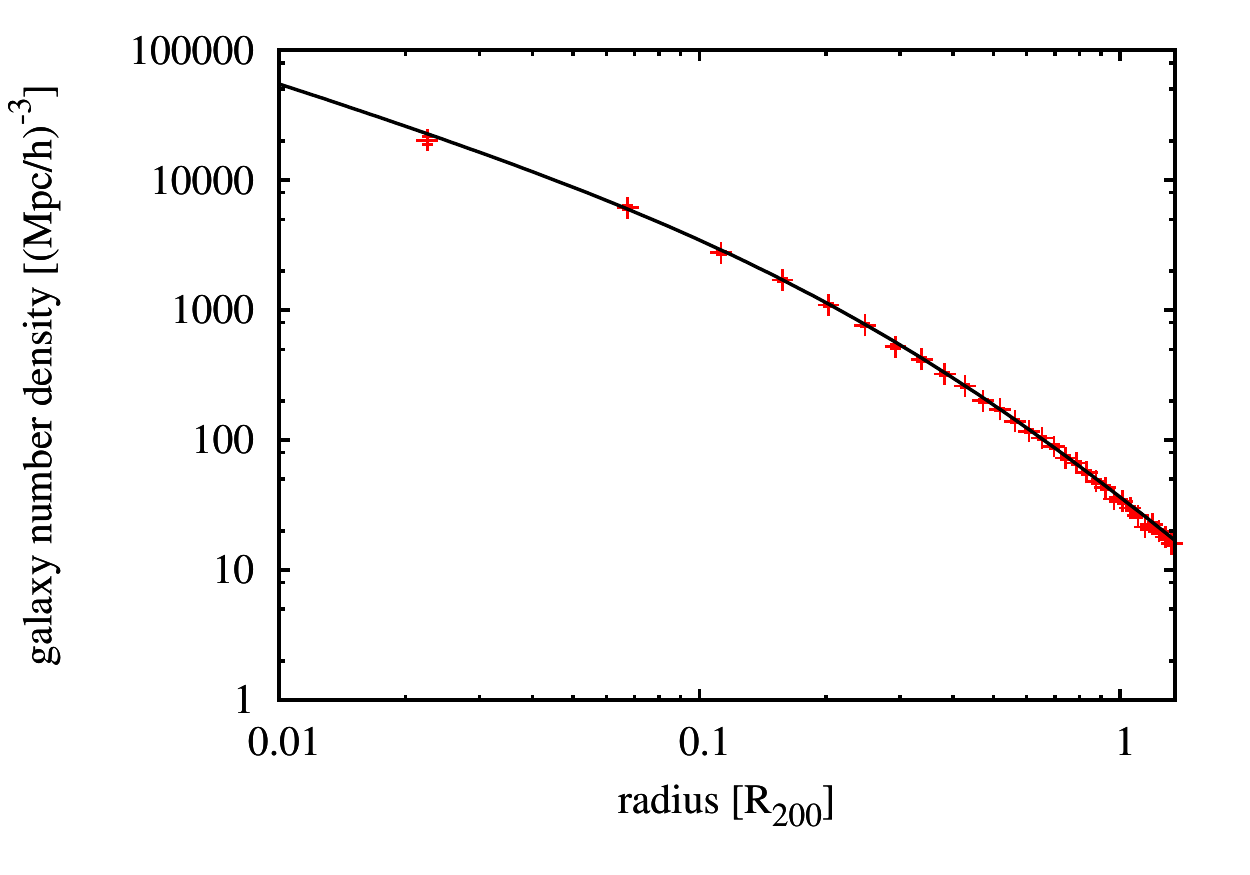}
 \includegraphics[width=\columnwidth]{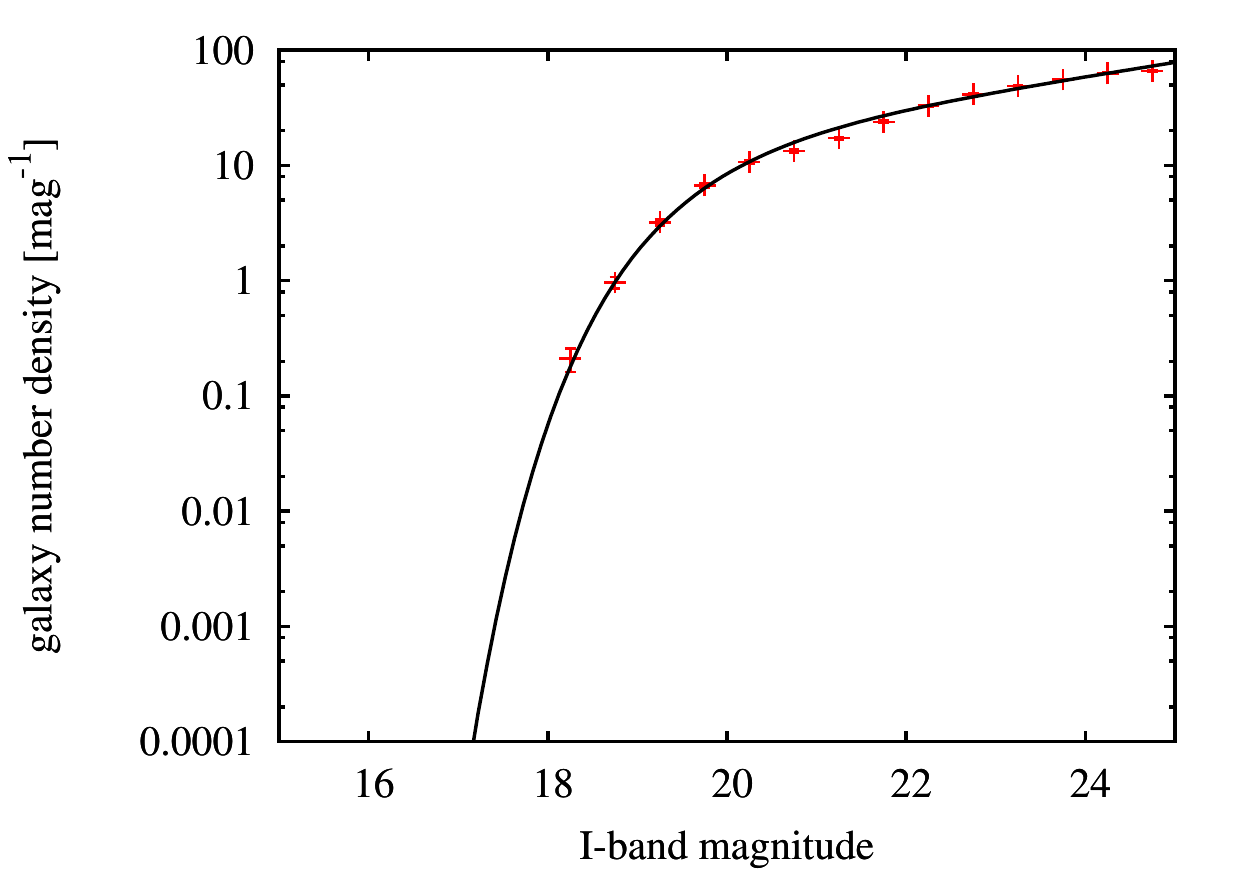}
 \caption{Empirical distributions and analytical fit for the cluster satellites at redshift $z$ = 0.4. Top panel: radial profile. Bottom panel: magnitude distribution. The error bars derive from Poissonian statistics, but are often smaller than the point size. These distributions were used to create the idealised mock simulations.}
 \label{fig:model_fit}
\end{figure}

As an example, we show the empirical distribution and the analytical fitting functions at $z$ = 0.4 in Figure \ref{fig:model_fit}. At all redshifts, the mean radial profile is consistent with a concentration $c$ = 3, so we use this value for the model. As for the other parameters (the total number of galaxies $N$, the typical magnitude $m_\star$, the faint-end slope $\alpha$), we perform a fit over redshift to get a smoothly changing model. Central galaxies are added separately, following a Gaussian distribution in magnitude and adding one galaxy in the first bin of the radial profile, as they lie by definition exactly at the center of the structures. This mimics the presence of a BCG in our model. For each redshift $z_c$, the desired $M_c(r,m)$ is then obtained by projecting the 3D radial distribution in 2D and multiplying it by the magnitude distribution.

The background distribution $N(m,z)$ is instead built from the galaxies which are not closer than $1.35R_{200}$ to any structure with $M_{200} > 0.5 \times10^{14} M_{\odot}/h$. This choice is meant to be complementary to the one made in constructing the model $M_c(r,m)$, following the separation in two parts of the galaxy population which is an assumption of the algorithm. Anyway, $N(m,z)$ is very weakly affected by the specific values of the thresholds in mass and radius, as the vast majority of the galaxy population resides in much smaller haloes.

Given the properties of the catalogue, we decided to concentrate our analysis over the redshift range $0.3 < z < 1$. In fact, at lower redshifts, the catalogue is not complete at the faint magnitudes, while at higher redshifts the galaxy density is so low that one cannot build properly the typical cluster distribution. In the remaining of the paper, we will restrict our analysis to this redshift interval. We note that, even if this work is not aimed at providing forecasts for any specific survey, this redshift range is common for cluster detection with optical data. AMICO can use the information coming from multi-band observations, but in the tests presented here we use only the $I$ band for simplicity.

 \subsection{Test on isolated clusters}\label{sect:big}
\begin{figure}
 \includegraphics[width=\columnwidth]{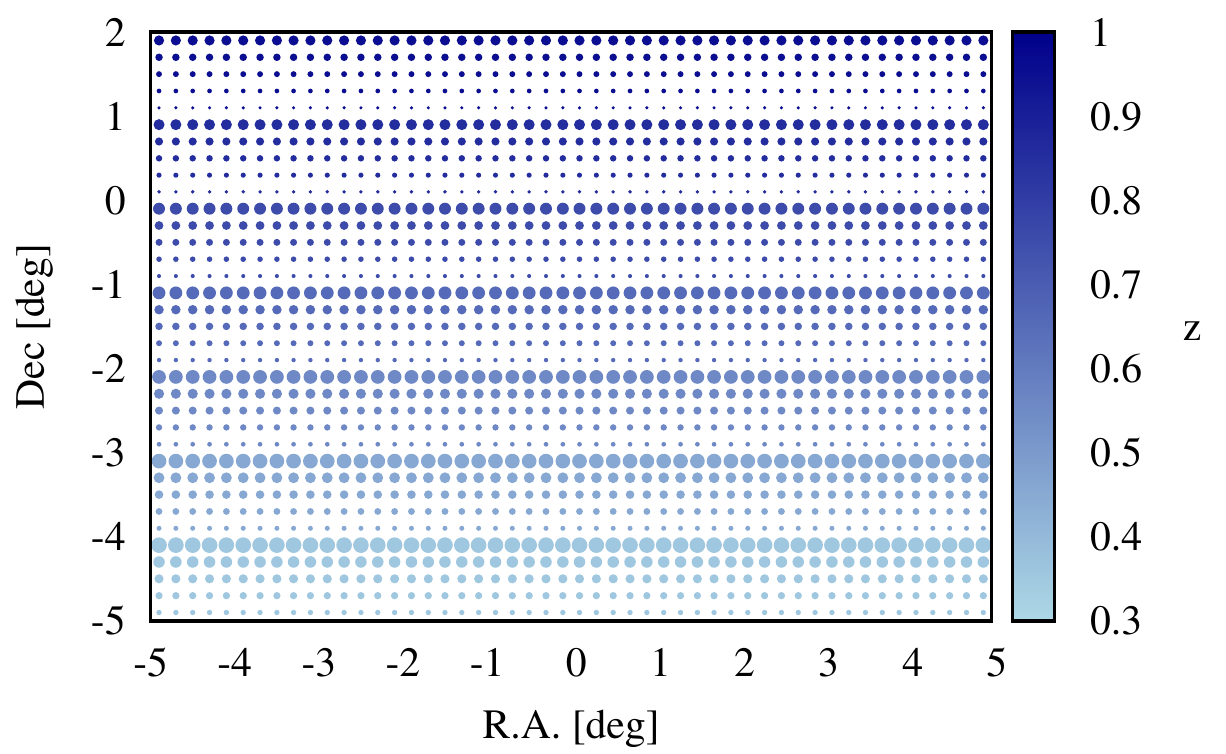}
 \caption{Distribution of clusters in the field described in Section \ref{sect:big}. The color indicates the input redshift, as coded by the bar on the right, the radius of the circle is proportional to the cubic root of the number of visible galaxies.}
 \label{fig:big_map}
\end{figure}

Having built the cluster model and the background component in the redshift range of interest, we can prepare our \textit{ideal} mock catalogues. As a first test case, we prepare a catalogue with structures disposed on a grid over a statistically uniform background. We consider seven redshift values, with $z$ ranging from 0.35 to 0.95, separated by $\Delta z$ = 0.1. For each redshift, we create 250 \textit{Monte-Carlo} realisations of our cluster model, 50 for each 5 different amplitude values $A_\text{true}$ = (0.2, 0.5, 1., 2., 5.). Clusters are distributed over a field with size $10 \times 7 \deg^2$ on a grid with a distance $0.2 \deg$ one from each other, as shown in Figure \ref{fig:big_map}. We underline that, following our assumptions, for a given redshift the expected galaxy distribution for clusters with any richness is the same, apart from a rescaling of the total number of galaxies. We add a homogeneous background component, following the density given by $N(m,z)$ and random positions over the field. Here and in the following, we applied a scatter to the redshift of the galaxies following a Gaussian distribution with $\sigma$ = 0.05(1+$z$), which is typical for wide-band photometric surveys. 

\begin{figure}
 \includegraphics[width=\columnwidth]{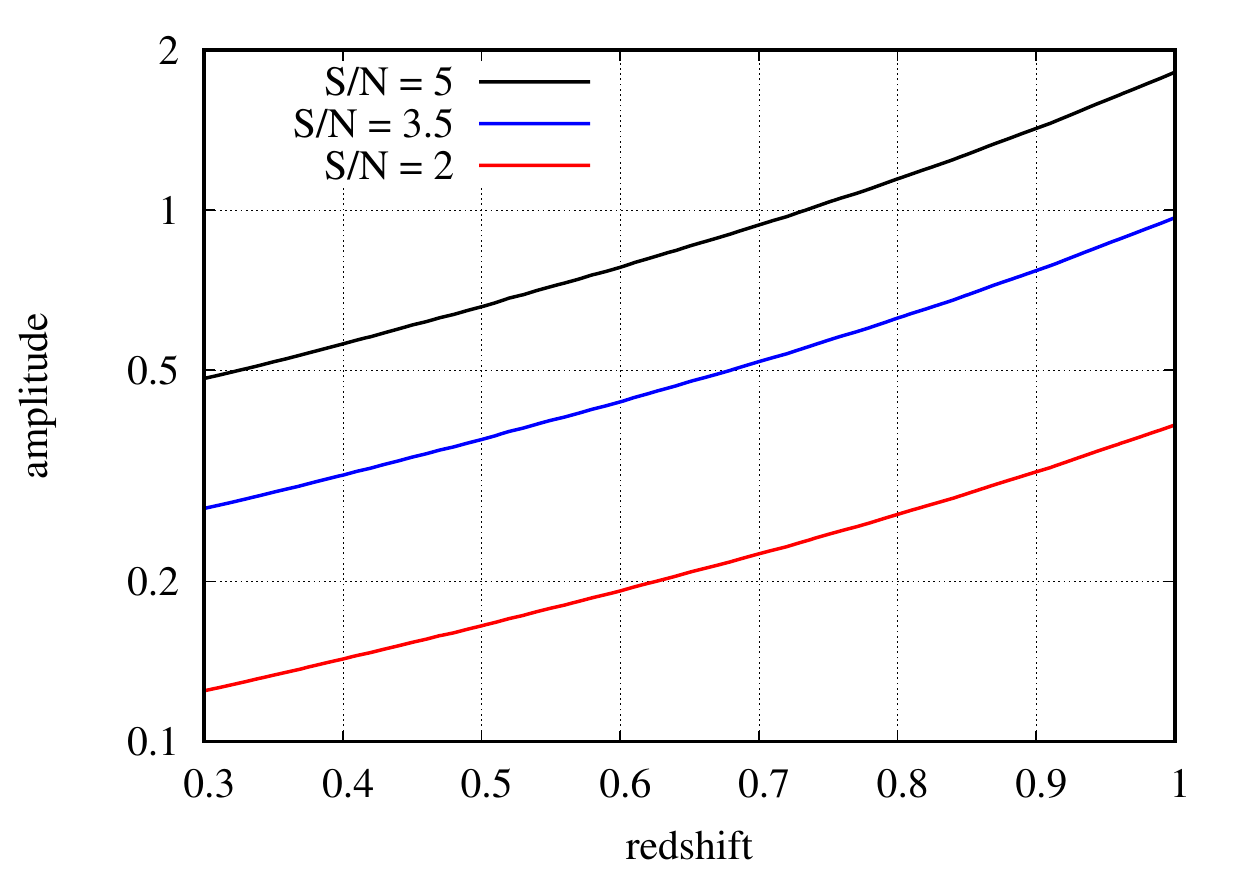}
 \caption{Theoretical minimum amplitude for detection of galaxy clusters as a function of redshift, for three different S/N thresholds.}
 \label{fig:analytic_sn}
\end{figure}

Given a cluster model and a background distribution, and having fixed the photometric redshift uncertainty, we can use Eqs. \ref{eq:real_amplitude}, \ref{eq:rough_amplitude} and \ref{eq:uncertainty} to estimate the expected signal-to-noise ratio for a cluster of any amplitude at any redshift, and build the expected selection function for our cluster search. The result is shown in Figure \ref{fig:analytic_sn}. This is a purely theoretical estimate that depends on our modellisation of the cluster and of the background and allows us to make a first estimate of the kind of objects we can detect under certain survey conditions. In particular, for any redshift we can compute the minimum amplitude $A_\text{min}$ required for the detection. In the following, we will compare this with the actual detection thresholds in our experiments on ideal and cosmological fields.
 
\subsubsection{Completeness}\label{sect:big_compl}

Once the mock catalogue is produced, we run AMICO on it using the model and the background estimated as above. In this case, as the background distribution is following our assumption of uniformity over the field, we do not use the local estimation presented in Section \ref{sect:localback}. The following quantities are read for each galaxy of the catalogue: angular position in the sky, magnitude in the $I$ band, photometric redshift estimate, photometric redshift uncertainty. The algorithm produces a list of detections as described in Section \ref{sect:selecting}. In order to be as complete as possible we set a S/N threshold = 2 even if, as we will see, the amount of spurious detections due to noise fluctuations is not negligible for S/N < 3. Once we have a list of detections, we try to match them with the actual structures in the field, allowing for an angular separation equal to the radius $R_{200}$ of each real structure and for a redshift separation equal to the 1-sigma uncertainty $0.05\times(1+z)$. When not explicitly mentioned, this \textit{matching procedure} will be repeated for all the experiments we describe.

\begin{figure}
 \includegraphics[width=\columnwidth]{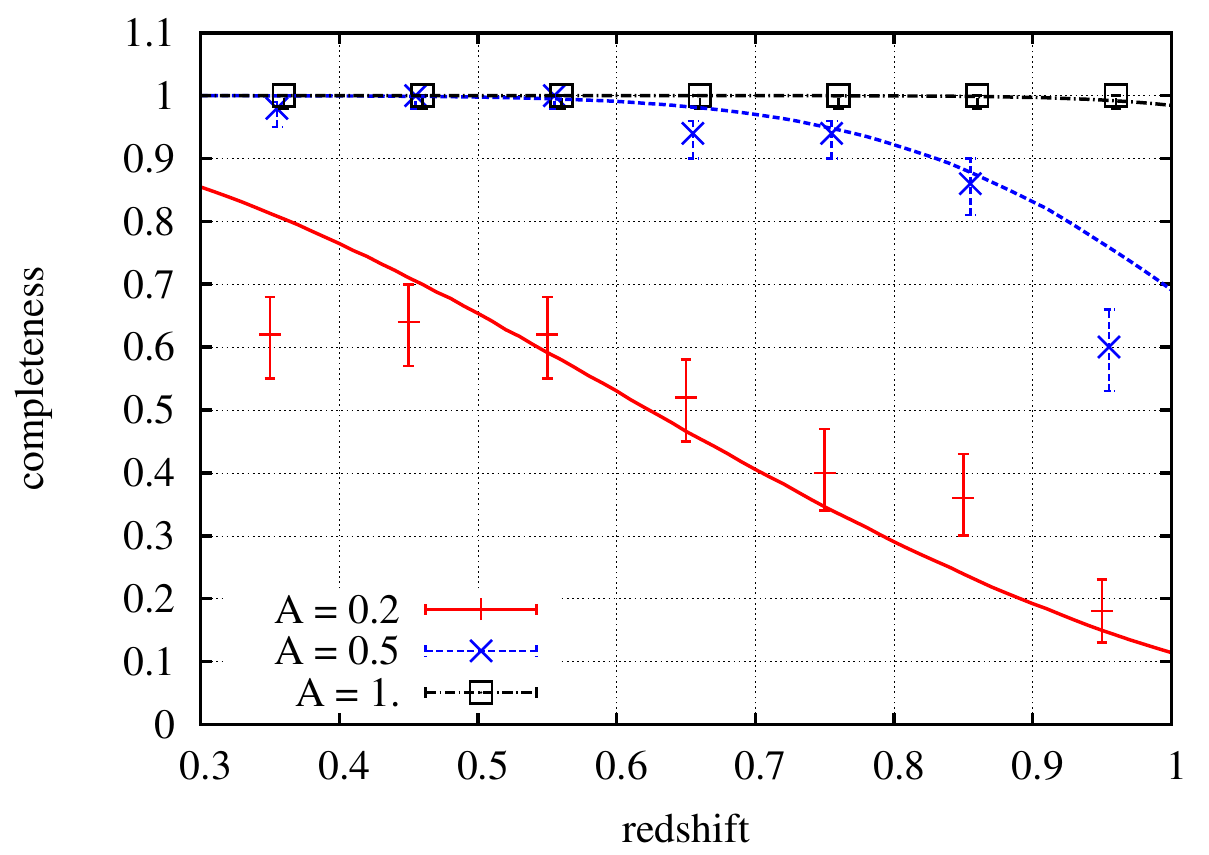}
 \caption{Points with error bars show the fraction of detected objects for different input amplitudes as a function of redshift. The lines indicate the theoretical detection probability. For clarity, points that refer to different input amplitude values are slightly offset along the x-axis. }
 \label{fig:big_compl}
\end{figure}

The resulting completeness for the 3 lower input values of amplitude (0.2,0.5,1.0) is shown in Figure \ref{fig:big_compl}. For comparison, we show also the theoretical detection probability which we obtain from Figure \ref{fig:analytic_sn}, considering a Gaussian scatter for the observed amplitude with mean equal to $A_{\text{true}}$ and uncertainty $\sigma_A$ calculated as in Equation \ref{eq:uncertainty}. The fraction of the resulting distribution of $A_{\text{obs}}$ which is greater than $A_{\text{min}}$ for the same redshift is the expected probability of detecting a structure with those properties. As we see, there is a good agreement between the theoretical prediction and the actual fraction of detected objects. For simplicity we do not show the results for the structures with input amplitude equal to 2 and 5, as the completeness is equal to 1 both in theory and in practice all over the redshift range for these objects.

\subsubsection{Properties of detected structures}\label{sect:big_properties}
\begin{figure}
 \includegraphics[width=\columnwidth]{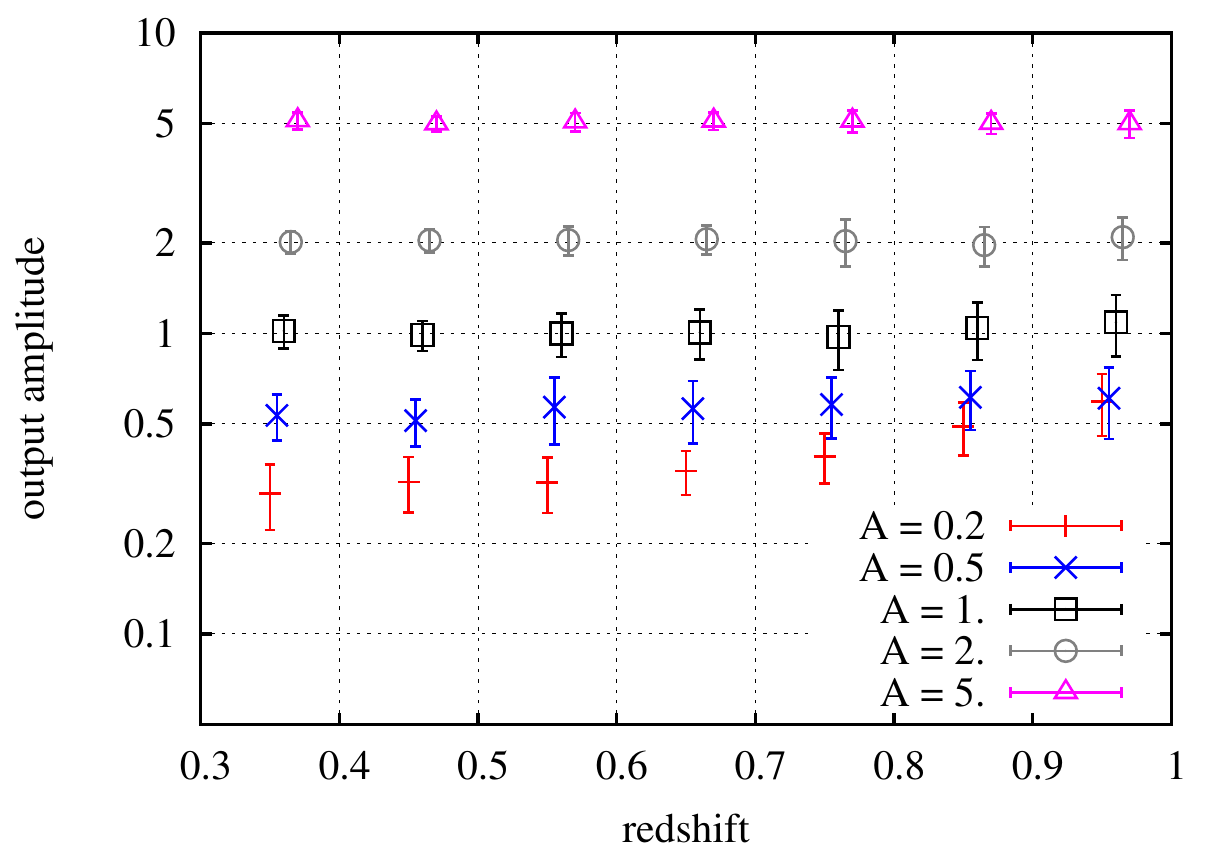}
 \caption{Mean and scatter of the measured amplitude as a function of the input redshift for structures with different input amplitudes. For clarity, points that refer to different input amplitude values are slightly offset along the x-axis.}
 \label{fig:big_rich}
\end{figure}

For the detected objects, we can compare the measured amplitude $A_\text{obs}$ with the input amplitude $A_\text{true}$. The results are shown in Figure \ref{fig:big_rich}. The estimated amplitude is unbiased with respect to the input value as long as the sample is complete, i.e. we are sampling the whole distribution of $A_\text{obs}$ for a given input $A_\text{true}$. This is true at all redshifts for the three most massive bins and at $z < 0.6$ for clusters with $A_\text{true}$ = 0.5 (see Figure \ref{fig:big_compl}). Instead, the sample of clusters with $A_\text{true}$ = 0.2 is never complete, because the theoretical amplitude of all objects lay below the detection line given by $A_{\text{true}}$, therefore only those objects whose $A_\text{obs}$ is increased by the random noise enter the detection catalogue, and this translates into a positive bias in the mean $A_\text{obs}$ for detected objects. The scatter in the measurement of $A$ is increasing for higher redshift and lower $A_\text{true}$, as expected given the S/N estimates.

\begin{figure}
 \includegraphics[width=\columnwidth]{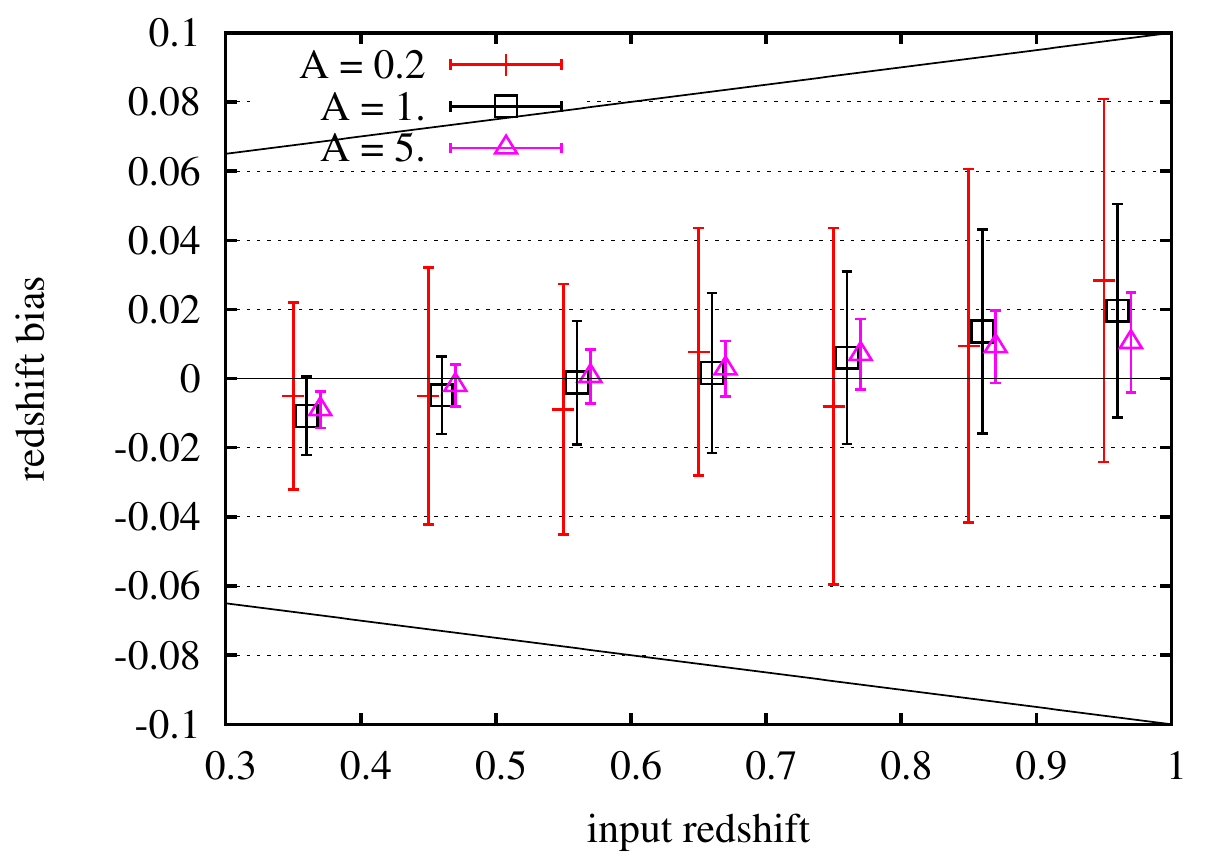}
 \caption{Mean and scatter of the redshift bias (output - input) as a function of the input redshift for structures with different input amplitudes. For clarity, the results for just 3 input amplitude values are shown and points that refer to different input amplitude values are slightly offset along the x-axis. The solid lines limit the region of 1-sigma uncertainty of the photometric redshift of individual galaxies. }
 \label{fig:big_red}
\end{figure}

The redshift of the detection can also be compared with the input one. As we see in Figure \ref{fig:big_red}, the redshift estimate is always consistent with the input value, considering the scatter. The error bars are shrinking for bigger structures, as the number of member galaxies increases and reduces the statistical uncertainty. We note that there is a slight tendency towards an under-estimation of $z$ for low redshifts and an over-estimation for high redshift, which is mildly significant for the biggest structures. This is due to the redshift evolution of the cluster model $M_c(r,m)$, that may cause the likelihood for a structure at redshift $z_c$ to peak at a slightly different redshift. As we will see in Section \ref{sect:real_mocks}, this source of bias is negligible in a realistic environment, with correlated noise and increasing confusion due to structures that may be close to each other and aligned along the line of sight. 

\subsubsection{Membership}\label{sect:big_members}
\begin{figure}
 \includegraphics[width=\columnwidth]{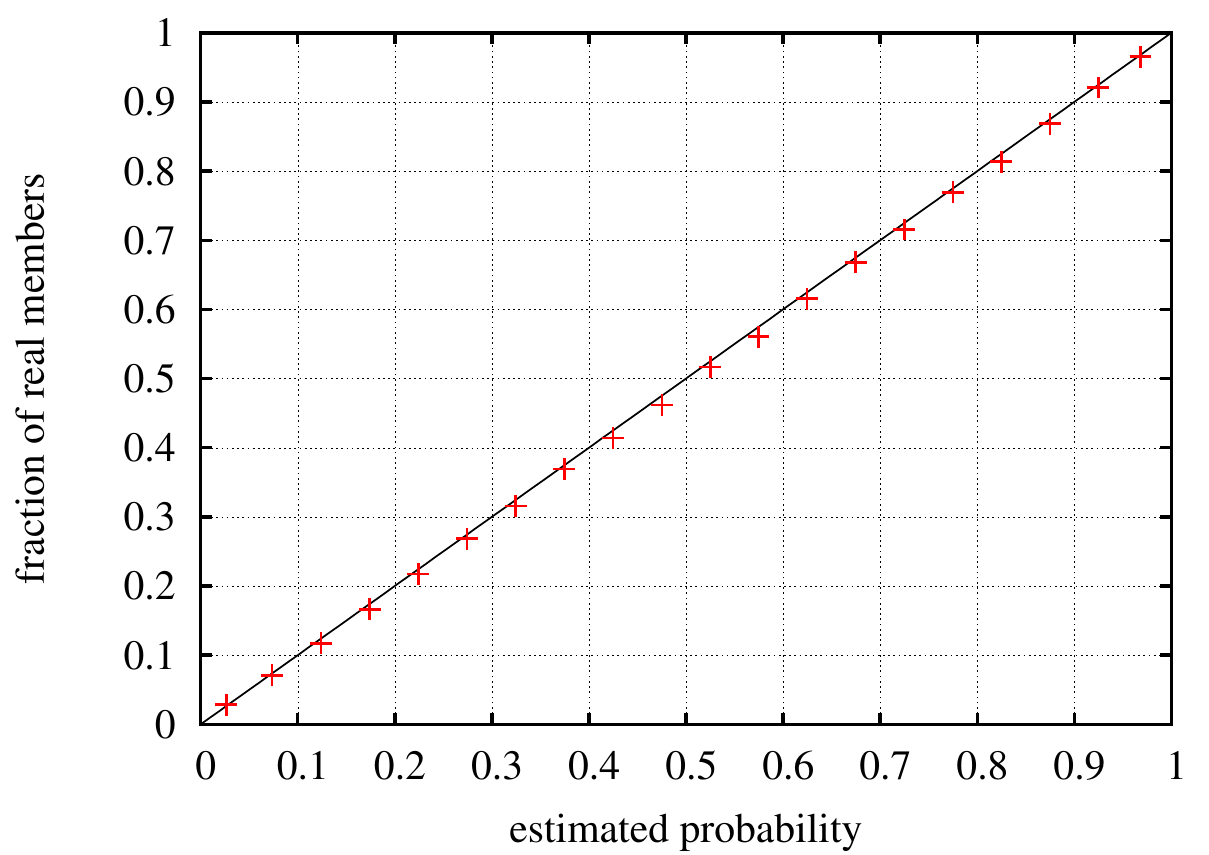}
 \caption{In red: mean membership probability as estimated by the algorithm versus the fraction of actual members, in bins of estimated probability. The error bars are negligible because they are smaller than the point size. The 1-to-1 relation is shown for reference as black solid line.}
 \label{fig:big_members}
\end{figure}

As described in Section \ref{sect:members}, after a detection is made, AMICO runs through the galaxy catalogue and assigns to each galaxy a probability of being a member of the identified object. We can verify how accurate is this probabilistic association by comparing it with the fraction of members that actually belong to the identified structure. The results are shown in Figure \ref{fig:big_members}. As we see there is an almost perfect agreement between the probability estimated by the code and the fraction of actual members. This is a very important result that verifies the effectiveness of the algorithm when all its assumptions are met. We will verify this result in a more realistic environment in Section \ref{sect:real_members}.

\subsubsection{Spurious detections}\label{sect:big_false}
\begin{figure}
 \includegraphics[width=\columnwidth]{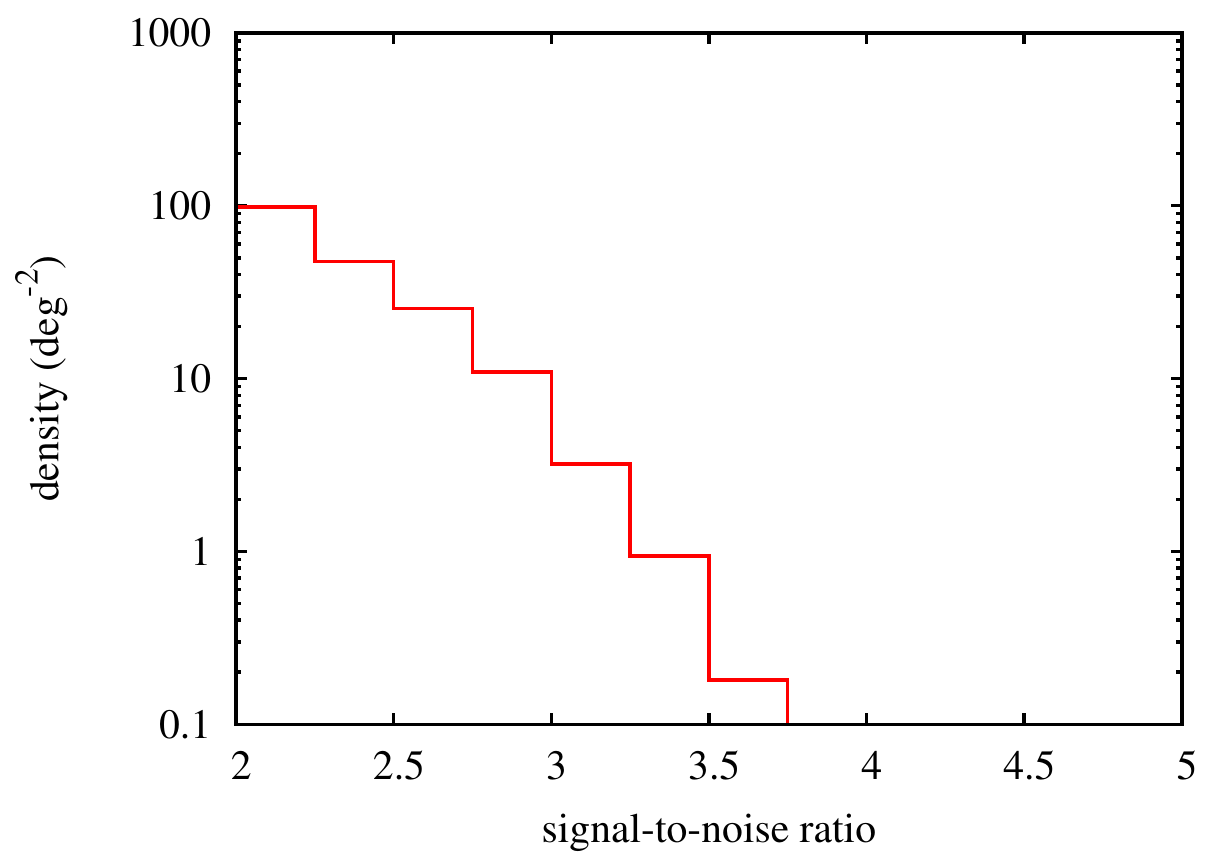}
 \caption{Number of spurious detections per square degree as a function of signal-to-noise ratio.}
 \label{fig:big_false}
\end{figure}

Up to now we focused only on detections which were matched to the objects we inserted in the catalogue. We now consider the fraction of detections that does not correspond to any structure, but instead appears just due to random fluctuations of the background distribution. Even if the Optimal Filtering formalism minimises their presence, their appearance is expected because of the probabilistic Gaussian nature of the measured amplitude which has the well defined variance given by Eq. \ref{eq:uncertainty}. Their distribution as a function of the signal to noise is shown in Figure \ref{fig:big_false}. As we see, the density of spurious detections decreases steeply with the signal-to-noise. We can estimate a density $\sim 4.5 \deg^{-2}$ at S/N > 3 and $\sim 0.2 \deg^{-2}$ at S/N > 3.5, which is a strong threshold against detections created by purely random background distribution. 

\subsection{Test on angular deblending }\label{sect:deblending}
In order to quantify the capability of AMICO to clean the galaxy catalogue from the detected structures (see Section \ref{sect:cleaning}) and allow for the identification of other neighbouring structures, we create a specific simulation with couples of clusters which are close to each other in angular distance (the same problem along the $z$ direction will be analysed in Section \ref{sect:cleaning_z}). The employed background distribution and cluster galaxy population are the same described in Section \ref{sect:big}. For this test, all the structures are at redshift 0.5 and have amplitude equal to 1. We created couples of clusters with a relative distance that goes from 0.01 to 0.1 $\deg$ in steps of 0.01 $\deg$. For each distance bin 25 couples were used.  

As it can be seen from Figure \ref{fig:analytic_sn}, the expected S/N for a single structure with the parameters we chose is larger than 5. For this reason, we expect that for every couple at least one of the two structures should be correctly identified, and this is verified. To verify how efficient is AMICO to disentangle the two components of the couples, we evaluate the completeness which here is defined as the fraction of couples which are correctly identified as two separate objects over the total number of pairs. 

\begin{figure}
 \includegraphics[width=\columnwidth]{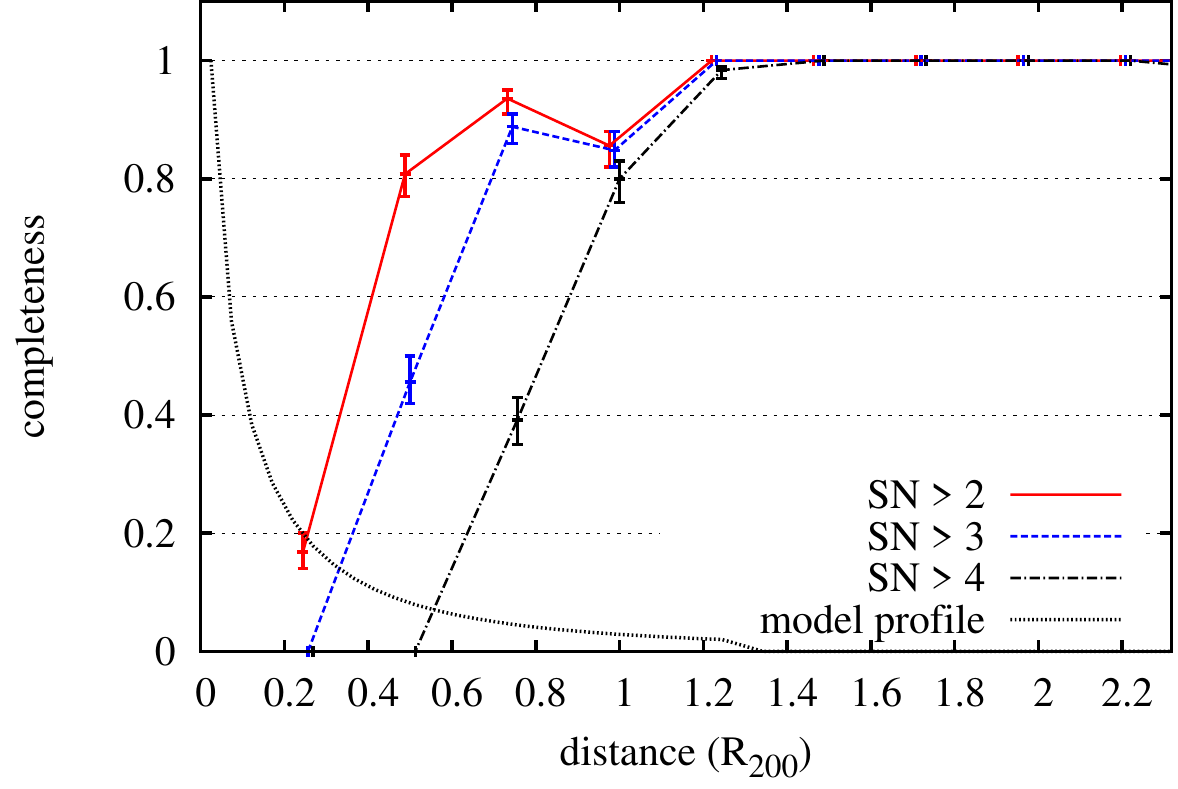}
 \caption{Points with error bars show the fraction of couples correctly identified as two separate objects as a function of the relative distance in degrees for different S/N thresholds (red: S/N $\geq$ 2, blue: S/N $\geq$ 3, black: S/N  $\geq$ 4). For reference, the radial profile of the cluster is shown as a dashed line.}
 \label{fig:cleaning_compl}
\end{figure}

\begin{figure}
 \includegraphics[width=\columnwidth]{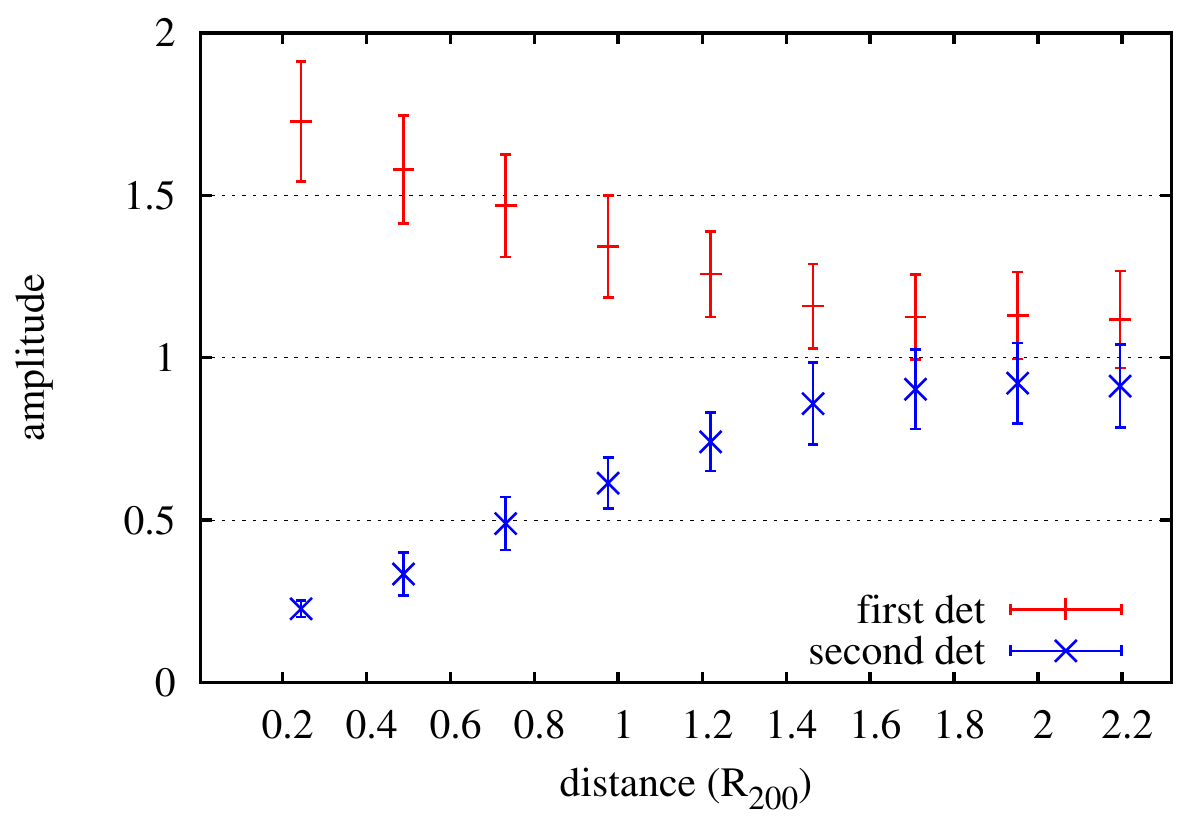}
 \caption{Mean and scatter of the estimated amplitude for the first (red) and the second (blue) detection of each couple as a function of the relative distance in degrees. Only couples for which both objects were detected are considered in this plot.}
 \label{fig:cleaning_amp}
\end{figure}

The results are displayed in Figure \ref{fig:cleaning_compl} for different thresholds in S/N, along with the projected profile of the cluster model. For the case under investigation, the deblending is virtually impossible for clusters with a relative distance = 0.01 $\deg$, while it becomes more and more likely for larger distances, and almost certain once the two structures are at a distance larger than $0.04 \deg$, which is close to their $R_{200}$. The completeness for different S/N thresholds highlights the fact that for close-by companions, the significance of the second detection is lowered with respect to the first one. This happens because, during the iterative detection procedure, the membership probability $P( i\in j )$ is assigned considering one detection at a time. Galaxies belonging to  the two objects concur to increase the amplitude estimation of the first detection, and the membership probability is then assigned following Equation \ref{eq:member_prob} to the galaxies surrounding it, including the ones that belong to the other object. For this reason, at the stage when AMICO finds the second detection the value of $P_{f,i}$ of its members is already $<1$, causing an underestimate in the resulting $A_{\rm obs}$ (see Equation \ref{amplitude_corr}). This can be seen also in Figure \ref{fig:cleaning_amp}, where the mean estimated amplitude for the first and the second detection of the couple are shown. In the filtering procedure the two structures interact with each other and, the closer they are, the more the first one ``steals'' amplitude from the second one.

That said, it is notable that AMICO with its cleaning approach is able to identify objects whose galaxy distributions are significantly overlapping on the sky and whose relative distance in some cases is smaller than $R_{200}$. In our simulations the two objects are exactly at the same redshift, but the same would apply to structures with relative line-of-sight distance lower than the typical photometric redshift uncertainty. We underline that the case we analysed is the most pessimistic from two points of view: the two structures have exactly the same redshift and the same input amplitude. Clearly, deblending of structures at different $z$ would be helped by their different photometric redshift distributions. Moreover, if one of the two structures is smaller than the other, the bias induced on $A_\text{obs}$ of the first detected structure would be less significant than the one shown in Figure \ref{fig:cleaning_amp}, making more accurate the probability association of galaxies which is needed in the following step of the iteration.

\subsection{Test on line-of-sight projections}\label{sect:cleaning_z}
Complementary to Section \ref{sect:deblending}, here we aim to study the possibility of AMICO to disentangle different structures when they are aligned along the line of sight but at different redshifts. To this end, we create a different simulation where clusters are made in couples. For each pair, the closest (foreground) object lies at $z$ = 0.5, while the second (background) one is at a redshift ranging from 0.54 to 0.9, in steps of 0.04. All the structures have an input amplitude equal to 1, but this corresponds to an expected S/N which is decreasing in redshift, as we can see from Figure \ref{fig:analytic_sn}. As in Section \ref{sect:deblending}, we are interested in the capability of the algorithm to detect both objects, as one of the two (in this case, the foreground one) is obviously identified. 

\begin{figure}
 \includegraphics[width=\columnwidth]{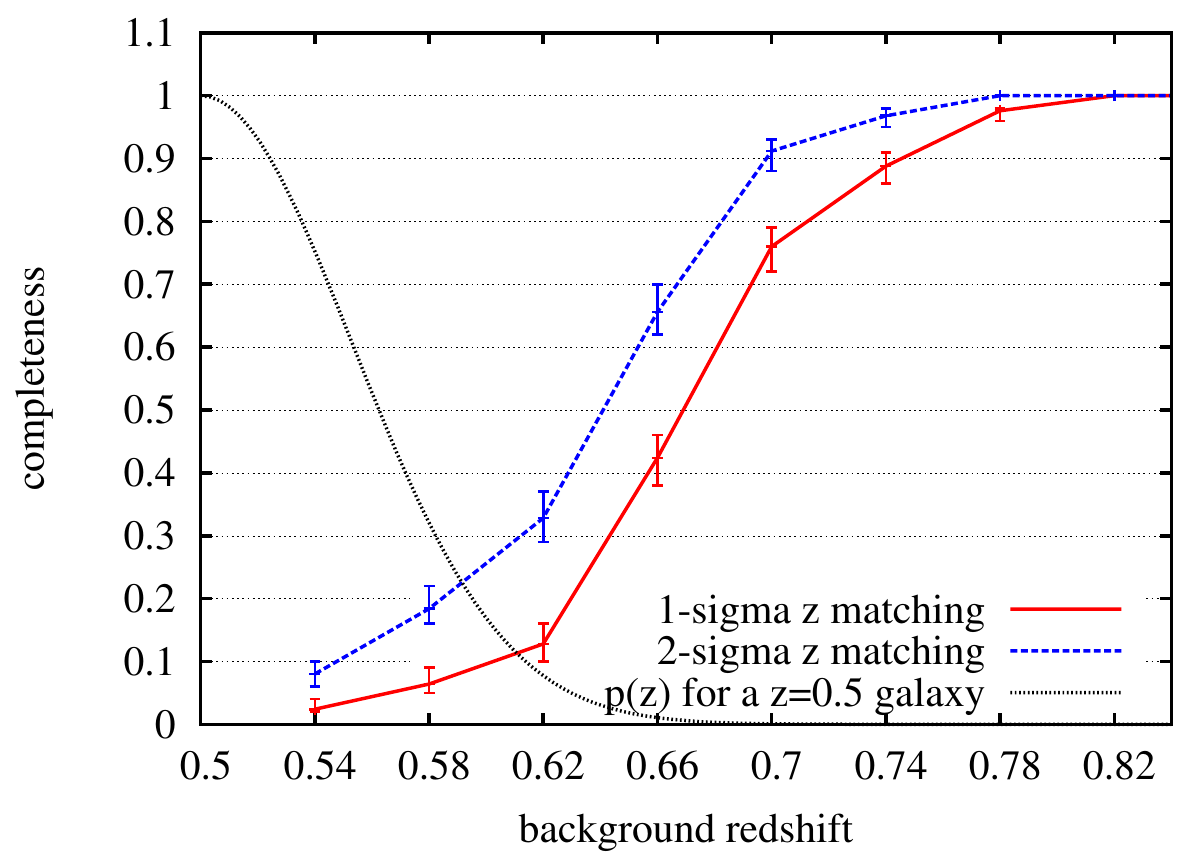}
 \caption{Points with error bars show the fraction of couples correctly identified as two separate objects as a function of the redshift of the most distant structure, for two different matching criteria in redshift: 1-sigma in red, 2-sigma in blue. The black dashed line is the shape of the photometric redshift distribution for the object at $z$ = 0.5.}
 \label{fig:cleaning_z_compl}
\end{figure}

\begin{figure}
 \includegraphics[width=\columnwidth]{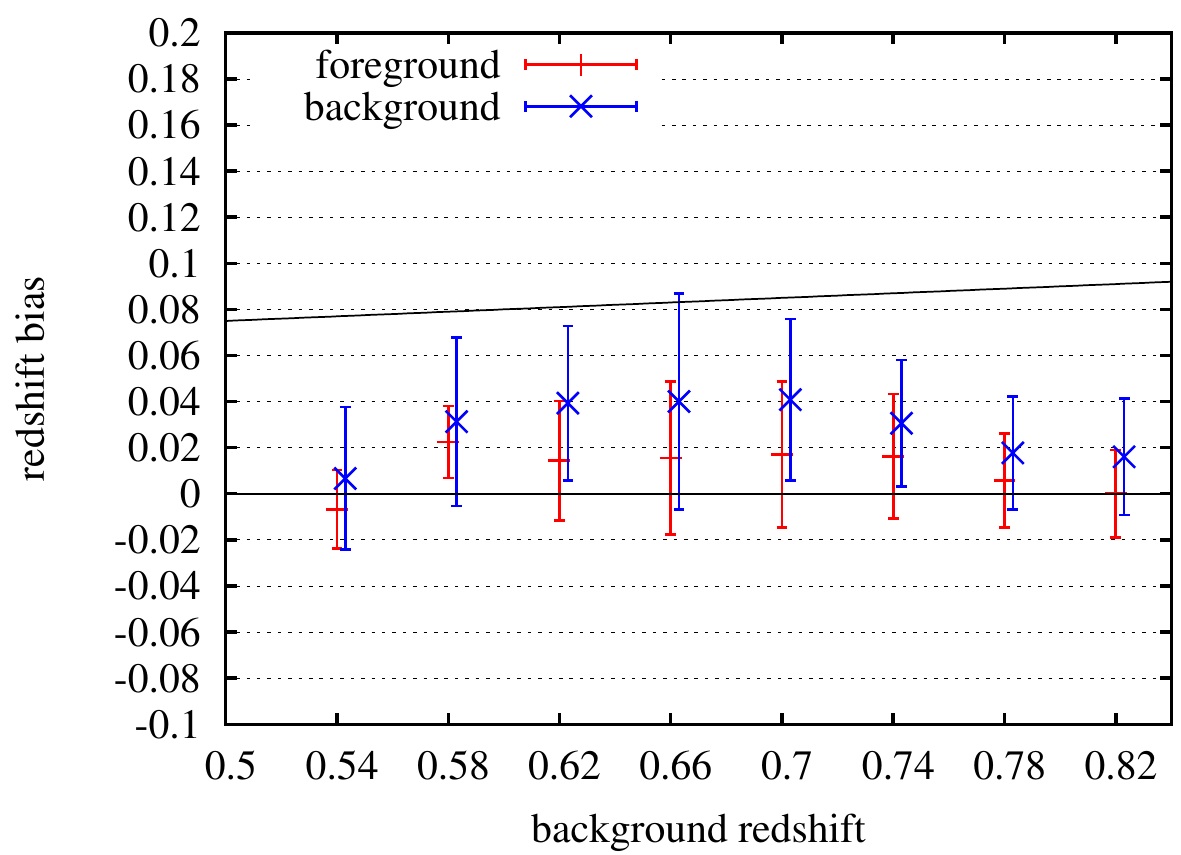}
 \includegraphics[width=\columnwidth]{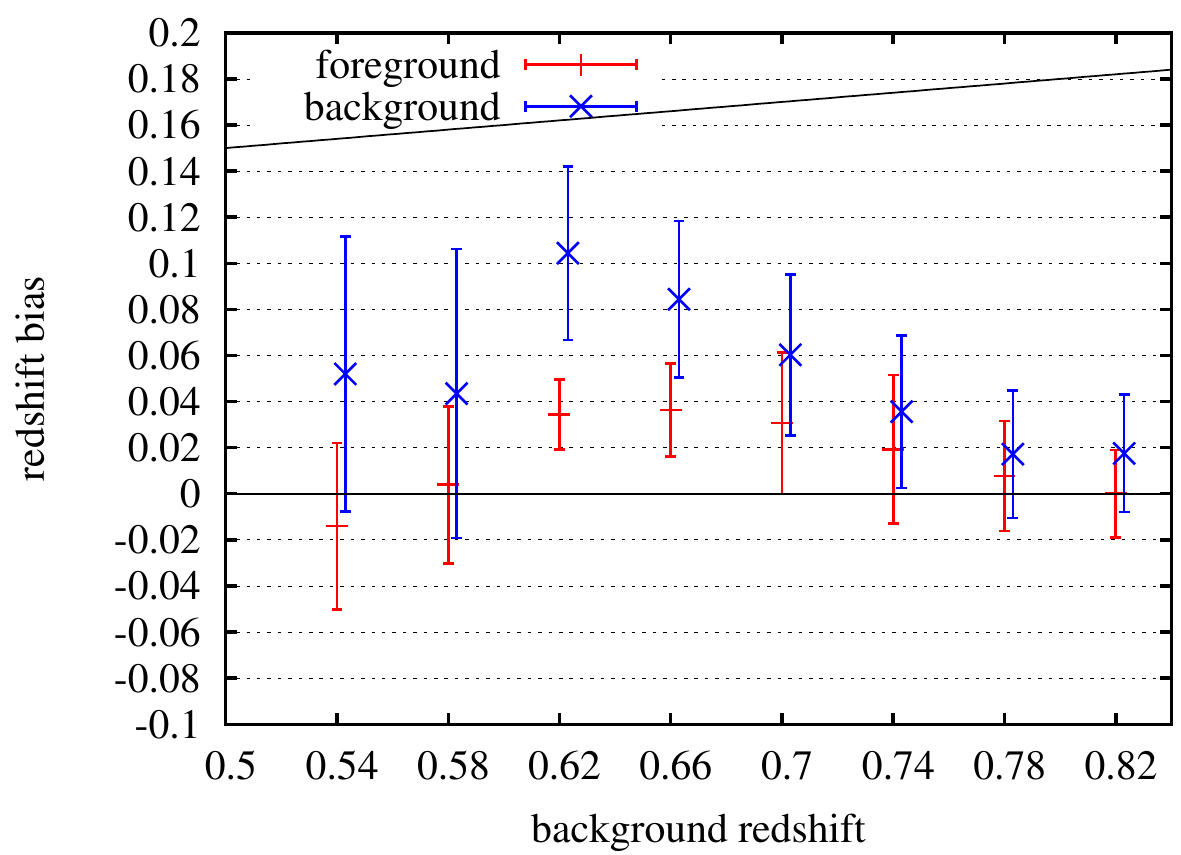}
 \caption{Mean and scatter of the bias in the estimated redshift for the first (red) and the second (blue) detection of each couple as a function of the redshift of the background structure. Top panel: 1-sigma $z$ matching. Bottom panel: 2-sigma $z$ matching. Only couples for which both objects were detected are considered in this plot. The upper solid line represents the limiting $z$ bias for which the object is considered matched in the two cases.}
 \label{fig:cleaning_z_red}
\end{figure}

The results as a function of the redshift of the background structure are shown as the red line in Figure \ref{fig:cleaning_z_compl}. We see that the fraction of detected structures goes above 50 \% when the relative separation in redshift $\Delta z_{pair}$ is larger than $0.16 \sim 2 \sigma_z$, while for smaller distances the confusion induced by the photometric redshifts makes it impossible to distinguish the two objects. 
We can better understand this behaviour by looking at the top panel in Figure \ref{fig:cleaning_z_red}, where the bias in the redshift determination for the foreground and the background structures is shown. Galaxies of the background structure tend to increase the signal, and thus the likelihood, of a detection at redshifts slightly higher that the true one of the foreground object, thus biasing high the measured redshift. In turn, the cleaning procedure for the foreground detection tends to remove galaxies of the other object which were scattered low in $z$. As a result, the redshift of the second detection is biased up as well. This phenomenon is especially relevant for $\sigma_z \leq \Delta z_{pair} \leq 3\sigma_z$, when the galaxies of the structures interact in the amplitude measurement. In particular, it seems likely that for $0.58 \leq z_\text{back} \leq 0.7$ some of the structures in the background have been detected but were not associated to a halo in the simulation because of the excessive redshift bias. In fact, as it is shown in Figure \ref{fig:cleaning_z_red}, in this redshift range the upper limit of the scatter of the redshift bias is close to or larger than the limit in $z$ for which the background object is considered matched, which is $0.05 \times (1+z)$.

To test this hypothesis, we relaxed the redshift criterion for matching to 2-sigma, equivalent to $0.1 \times (1+z)$, and the result for the completeness is the blue line in Figure \ref{fig:cleaning_z_compl}. We see there is an increase in the fraction of matched couples, and this is especially relevant for $z_\text{back} = 0.62$ and $z_\text{back} = 0.66$ whose completeness increases by $\sim 20 \%$. The corresponding plot about redshift measurement is in the bottom panel of Figure \ref{fig:cleaning_z_red}. We see the increase in the redshift bias especially for $z_\text{back} = 0.62$ and $z_\text{back} = 0.66$, due to the insertion in the catalogue of couples whose redshift is significantly biased high, confirming our hypothesis. Now, the distribution of measured redshifts is well distant from the matching limit, indicating that its exact location is not affecting anymore the matching results.

Again, we note that these results are relative to a very pessimistic case (two structures at equal amplitude perfectly aligned on the sky) and thus must be considered as a worst-case scenario when facing this problem. We note anyway that AMICO is able in a relevant fraction of cases to detect a structure which lies just behind another one, completely overlapping in the sky and with the two galaxy populations which are overlapping in redshift, with a $\Delta z \sim 2 \times \sigma_z$, where $\sigma_z$ is the typical photometric redshift uncertainty of the galaxies.

As a general remark, the cases presented here and in Section \ref{sect:deblending} quantify the possible biases in the determination of clusters position and amplitude when two or more structures are nearby, and the description of the data as the sum of a field and a cluster component fails. A simple check on the spatial proximity of the output detections of AMICO provides a list of possible cases where these small systematics may occur.

\section{Tests on cosmological mocks}\label{sect:real_mocks}
Now we turn to the analysis of the mock catalogues presented in Section \ref{sect:simul}, as they are built from the N-body cosmological simulation by applying a semi-analytical modelling of the galaxy evolution. To make the original catalogues more realistic we introduced a scatter in the redshift of the galaxies to simulate the photometric redshift uncertainties expected from ground-based observations. As in the previous Section, galaxies were scattered following a Gaussian distribution with r.m.s. $\sigma_z = 0.05 \times (1+z)$. The 50 mock catalogues with a size equal to $2 \times 2 \deg^2$ were analysed separately and then the matching with real haloes was performed following the same procedure described in Section \ref{sect:big_compl}. We initially considered in the matching all the haloes with mass $M_{200}$ larger than $10^{12} M_\odot/h$ as identified in the catalogue. In Section \ref{sect:real_massrich} we will discuss and modify this mass threshold for the halo catalogue. Position and redshift of the haloes are set equal to the ones of their central galaxy, as indicated in the catalogue. As in the previous Section, the cluster search was performed in the redshift range $0.3 \leq z \leq 1$. We used the same model $M_c(r,m)$ and background distribution $N(m,z)$ we described in Section \ref{sect:model}. Differently from the analysis of the ideal mocks, in this case we ran AMICO applying the local background estimation (see Section \ref{sect:localback}) that helps taking into account the large-scale fluctuations of the field. We discuss its impact on the quality of the detections in Section \ref{sect:real_massrich}. 

\subsection{Completeness}
\begin{figure}
 \includegraphics[width=\columnwidth]{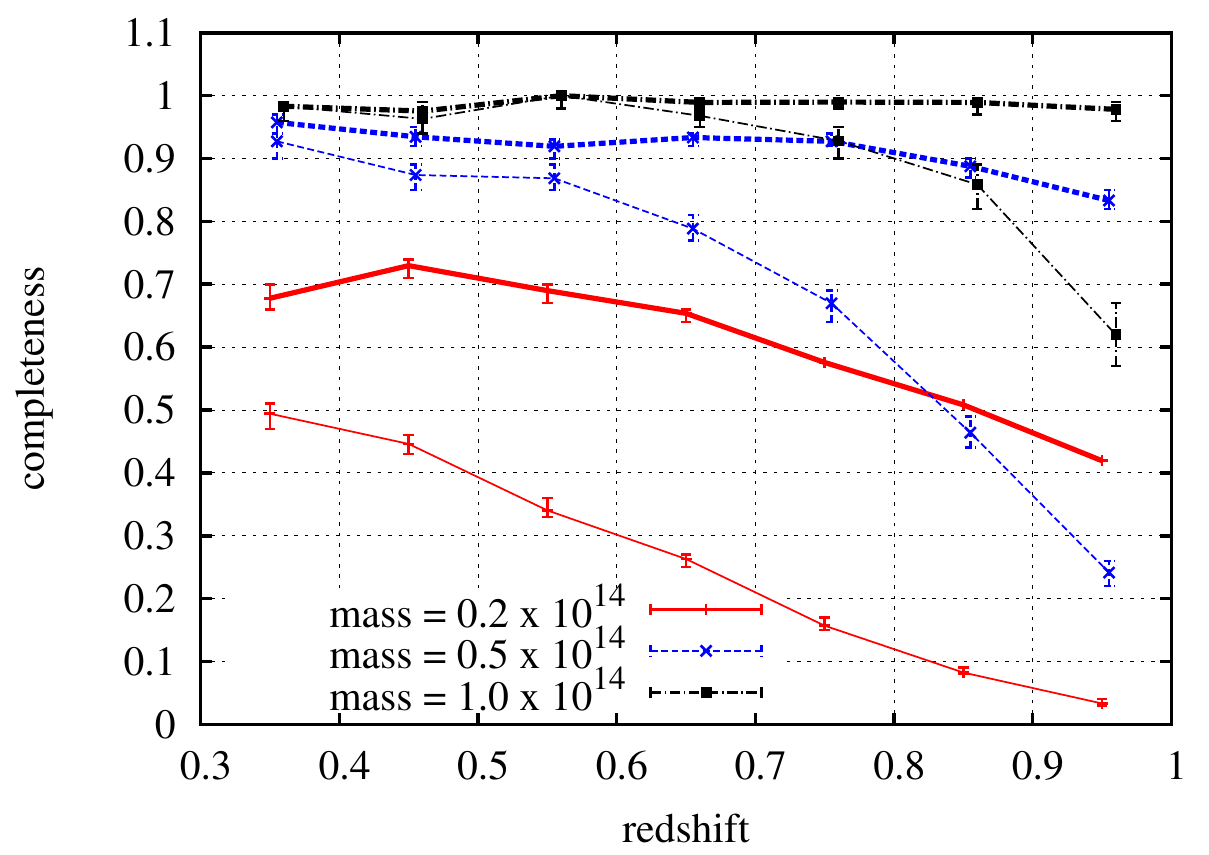}
 \caption{Fraction of detected structures (completeness) as a function of redshift for three different bins in logarithmic mass ($\Delta \log M = 0.1$) and two S/N thresholds: S/N > 2 in solid, S/N > 3.5 in dashed. }
 \label{fig:50_compl}
\end{figure}

Once the matching is performed, we can quantify the fraction of real objects detected by AMICO per bin of redshift and mass. We plot the results in Figure \ref{fig:50_compl}, where for simplicity we show only three mass bins, corresponding to masses respectively $\sim 0.2, \sim 0.5, \sim 1.0 \times 10^{14} M_\odot/h$, each with a size $\Delta \log M = 0.1$. The completeness is shown for two different S/N cuts: S/N > 2 as we used in Section \ref{sect:ideal}, and S/N > 3.5, which protects against pure random fluctuations of the background, as shown in Section \ref{sect:big_false}. In general, as one would expect, the completeness is decreasing both with redshift and with mass. 

Relating the measured completeness as a function of mass with the analytical estimates as a function of amplitude shown in Figure \ref{fig:analytic_sn} is not trivial, because of the mass-amplitude relation, which we will analyse in Section \ref{sect:real_massrich}. The only sample for which we can directly compare results with the analytical expectations is the one with mass $\sim 1.0 \times 10^{14} M_\odot/h$. Objects in this mass bin are detected with completeness close to 1 at S/N > 2 for the entire redshift range. The completeness at S/N > 3.5 of this sample is instead decreasing at high redshift, coherently with the theoretical S/N which is close to 3.5 at $z$ $\geq$ 0.9, as shown in Figure \ref{fig:analytic_sn}. 

\subsection{Redshift estimation}\label{sect:real_redshift}
\begin{figure}
 \includegraphics[width=\columnwidth]{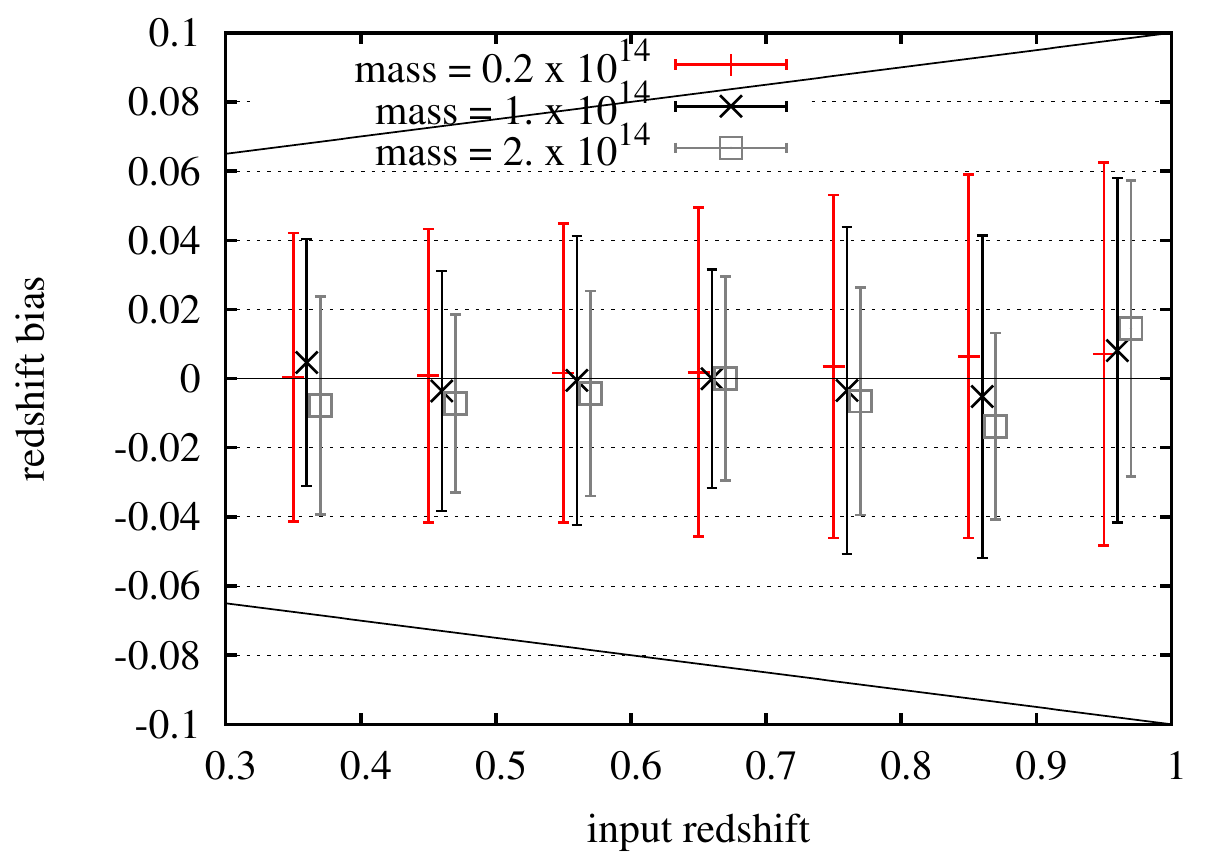}
 \caption{Mean and scatter of the redshift bias (output - input) as a function of the input redshift for structures with different masses. For clarity, just 3 mass bins are shown, each with $\Delta \log M = 0.1$, and points that refer to different mass bins are slightly offset along the x-axis. The solid lines limit the region of 1-sigma uncertainty of the photometric redshift.}
 \label{fig:50_red}
\end{figure}

We can perform in this more realistic environment the same analysis we performed in Section \ref{sect:big_properties} about the redshift estimation. The results are shown in Figure \ref{fig:50_red} for three different mass bins. Comparing the results with Figure \ref{fig:big_red}, we see that the scatter is not as much dependent on mass as it was on amplitude on the ideal mock. The small bias that was visible in the idealised situation is completely negligible in the realistic case. Both these differences depend on the presence of other sources of noise in these mock catalogues, such as the structures that are located along the line of sight to the detected objects, making the background much more complicated than our assumption of a random component with uniform mean. We see that the scatter is also mildly dependent on redshift, as expected since the scatter in the photometric redshifts is linearly dependent on (1+$z$), and is of the order of $0.5 \sigma_z$. This value may look large considering the number of cluster members which contribute to the redshift determination, but the precision is limited by the intervening galaxies belonging to the large-scale structure. The absence of redshift bias in this test depends also on the choice of using an unbiased Gaussian distribution for the redshifts of the galaxies in the mocks. In applications to real data, some bias in the cluster redshift might arise if photometric redshifts in the catalogue are biased.

\subsection{Mass-amplitude relation}\label{sect:real_massrich}
\begin{figure}
 \includegraphics[width=\columnwidth]{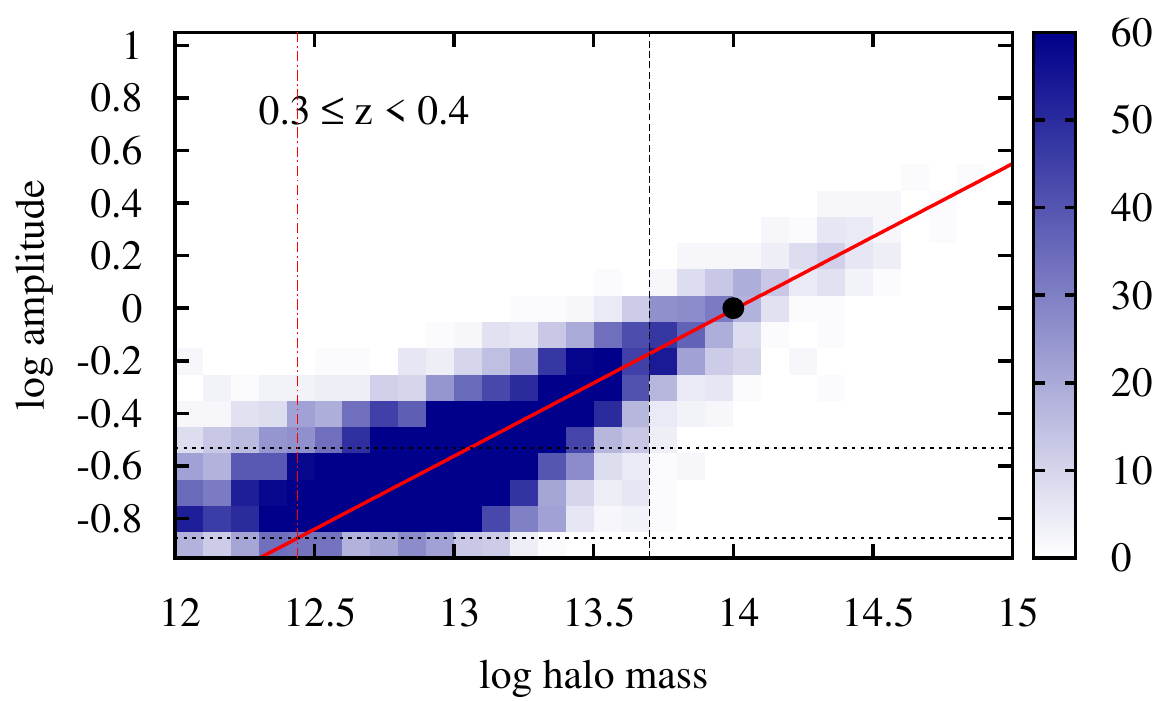}
 \includegraphics[width=\columnwidth]{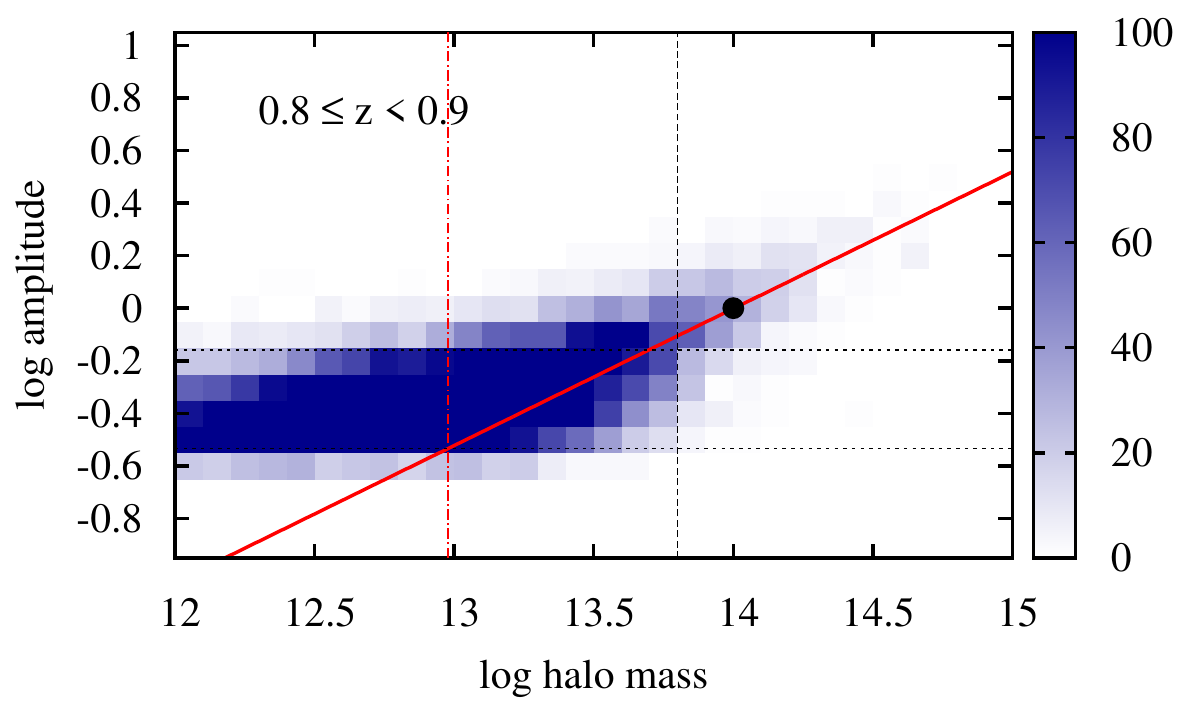}
 \caption{Density plot for the relation between halo mass and estimated amplitude for two redshift bins: 0.3 $\leq z <$ 0.4 in the top panel, 0.8 $\leq z <$ 0.9 in the bottom panel. The red line marks the best-fit mass-amplitude relation. The dashed black vertical line indicates the 90\% completeness limit for that redshift bin. Only objects with mass larger than that threshold were considered in the fit. The two horizontal lines indicate the amplitude corresponding to a S/N equal to 2 (threshold for detection) and 3.5. The dashed red vertical line indicates the minimum mass for which the expected amplitude corresponds to a S/N above the detection threshold, according to the best-fit mass-amplitude relation. The black dot indicates the original calibration of the model, such that amplitude equal to 1 corresponds to a mass of $10^{14} M_\odot/h$.}
 \label{fig:50_massrich}
\end{figure}

One of the most important properties of a detection method for galaxy clusters, especially for cosmological exploitation, is the relation between the halo mass and the observable. In the case of AMICO the mass proxy is provided by the amplitude $A$ of the detection (see Eq. \ref{eq:real_amplitude}.) The density plots for the mass-amplitude relation in two different redshift bins are shown in Figure \ref{fig:50_massrich}, together with the best-fit log-linear relation $\log A = a\times \log M + b$. To avoid a more complete treatment that takes into account the incompleteness of the sample and the intrinsic scatter \citep[see e.g.][]{2012A&A...547A.117A,2015MNRAS.450.3633S}, the best-fit relation is computed considering only the mass range for which the completeness is larger than 90\%, to have an almost complete sample of measured amplitudes for a given bin of mass and redshift. The lower limit of this mass range goes from $\log M/(M_\odot/h) = 13.6$ at low $z$ to $\log M/(M_\odot/h) = 13.9$ in the last redshift bin. As it can be seen in Table \ref{tab:mass-rich}, the slope of the relation is between 0.5 and 0.6 at any redshift, somewhat flatter than what is usually found for other richness estimators in literature \citep[see e.g.][]{2012ApJ...746..178R,2016A&A...586A..43V,2017MNRAS.468.3322S,2017MNRAS.466.3103S}. This may depend on the fact that the amplitude is computed by AMICO with a fixed filter, which does not change according to the physical size of the object. We expect that for objects less massive than $10^{14} M_\odot/h$ (and thus typically smaller than the model) the correlated noise due to large-scale structure increases the measured amplitude. On the other hand, for more massive and larger objects we miss part of the galaxy population that exceeds the model radius. It is in principle possible to extend the filter to the observed size of the objects, employing the scale-adaptive version of it \citep{2010MNRAS.403..859V}.

We point out the very precise estimate of the amplitude for objects with mass $\sim 10^{14} M_\odot/h$. In fact, as the model has been derived from the galaxy population of this sample of clusters, their expected amplitude is equal to one by construction. AMICO recovers perfectly this value, with bias $b$ consistent with zero over the whole redshift range, with no significant difference with respect to the \textit{ideal} case presented in Section \ref{sect:big_properties}. This proves that even when dealing with objects with different shapes, luminosity functions and embedded in the large-scale structure, AMICO provides an unbiased mass proxy, provided that the average cluster properties are known.

In Table \ref{tab:mass-rich} we also show the scatter of the amplitude at given mass $\sigma_{A|M}$, and the scatter of mass at given amplitude $\sigma_{M|A}$. The latter was estimated from structures detected at S/N $\ge$ 3.5. The shallow $A-M$ relation makes $\sigma_{M|A}$ much larger than $\sigma_{A|M}$. However, both these values are somewhat dependent on the sample definition, because they decrease for larger amplitude and mass.

Finally, we report in Table \ref{tab:mass-rich} the parameters of the mass-amplitude relation which are obtained assuming a spatially uniform background (i.e. neglecting the local background estimate described in Section \ref{sect:localback}). By comparing parameters with and without local background correction, we see that the advantage of applying the $f(\theta,z_c)$ local correction is twofold: the intercept $b$ of the relation is closer to zero and the scatter is reduced. This is consistent with the expectation that massive haloes are embedded in a higher background that, when neglected, biases high the amplitude estimate. It also proves the ability of our method to correctly capture these large-scale features, reducing their contribution to the $A-M$ scatter.

\begin{table*}
\centering	
\caption{Best-fit parameters for the relation $\log A = a\times \log M + b$ between mass $M$ and amplitude $A$ for different redshift bins. Between parentheses we show the same parameters of the relation when AMICO is run without local background correction.}
\label{tab:mass-rich}
\setlength\tabcolsep{0.8\tabcolsep}
\begin{tabular}{ccccc}
\hline
$z$ range & slope $a$ & intercept $b$ & scatter of $\log A$ & scatter of $\log M$ \\
$[0.3,0.4[$ & $0.556 \pm 0.045$ ($0.599 \pm 0.036$) & $-0.006 \pm 0.011$ ($0.028 \pm 0.010$) & 0.125 (0.156) & 0.290 (0.310)\\
$[0.4,0.5[$ & $0.580 \pm 0.032$ ($0.592 \pm 0.031$) & $-0.001 \pm 0.009$ ($0.030 \pm 0.009$) & 0.126 (0.150) & 0.300 (0.326)\\
$[0.5,0.6[$ & $0.563 \pm 0.036$ ($0.568 \pm 0.035$) & $0.000 \pm 0.009$ ($0.030 \pm 0.009$) & 0.140 (0.146) & 0.313 (0.344)\\
$[0.6,0.7[$ & $0.554 \pm 0.039$ ($0.563 \pm 0.031$) & $-0.009 \pm 0.009$ ($0.019 \pm 0.009$) & 0.135 (0.156) & 0.298 (0.326)\\
$[0.7,0.8[$ & $0.529 \pm 0.046$ ($0.540 \pm 0.045$) & $-0.011 \pm 0.010$ ($0.022 \pm 0.010$) & 0.144 (0.164) & 0.332 (0.343)\\
$[0.8,0.9[$ & $0.521 \pm 0.065$ ($0.531 \pm 0.049$) & $-0.001 \pm 0.013$ ($0.033 \pm 0.011$) & 0.151 (0.158) & 0.366 (0.395)\\
$[0.9,1.0[$ & $0.510 \pm 0.092$ ($0.492 \pm 0.068$) & $-0.008 \pm 0.020$ ($0.033 \pm 0.013$) & 0.137 (0.151) & 0.378 (0.392)\\

		\hline
	\end{tabular}
\end{table*}

\subsection{Miscentering}\label{sect:miscent}

\begin{figure}
 \includegraphics[width=\columnwidth]{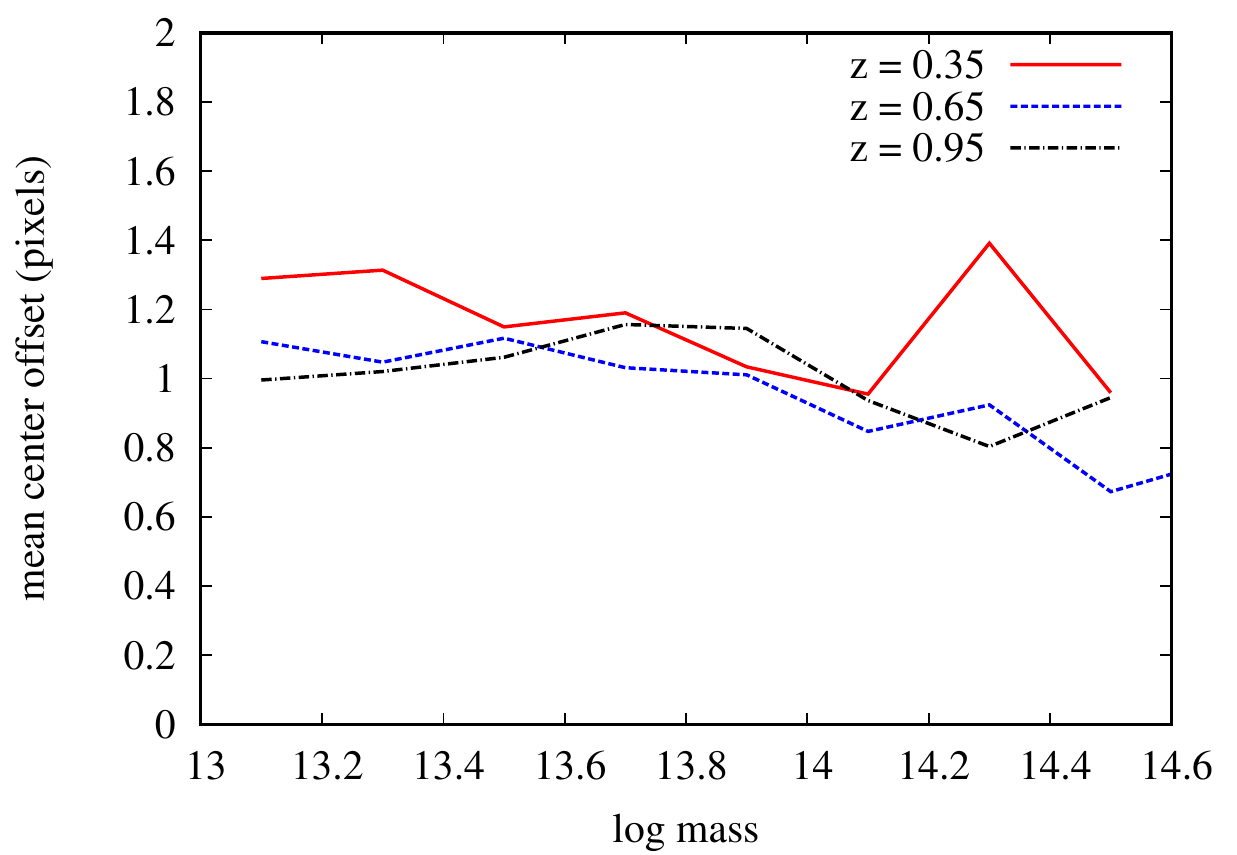}
 \caption{Mean offset between the center of the halo and the center of the detection, in pixels, as a function of halo mass. The pixel size is set to 0.005 deg which correspond to 63.7, 90.2 and 101.3 kpc/h at redshift 0.35, 0.65 and 0.95, respectively. }
 \label{fig:50_miscent}
\end{figure}

For several applications (e.g. weak lensing mass measurements, galaxy density profiles) it is important to estimate accurately the center of the detected structure. AMICO, differently from other methods, does not search for a central galaxy, but instead selects the center as the position of highest likelihood, that at a given redshift corresponds to a maximum in amplitude as well (See Equation \ref{eq:likelihood}). The identification of the central galaxy can then be done a posteriori according to the specific goal. We can measure the mean offset between the center as defined by AMICO and the central galaxy as indicated in the catalogue. The results are shown in Figure \ref{fig:50_miscent}. The mean offset is typically of the order of the size of the amplitude map pixel (0.005 deg). In physical units, this translates in a value between 60 and 120 kpc/h that tends to increase with redshift. In principle, the resolution of the amplitude map can be increased at will, but there is a trade-off between precision and computing speed. Moreover, given that our center depends on a filtering procedure, it is meaningless to give it with a precision smaller than the typical size over which the filter changes significantly.

We underline that in Figure \ref{fig:50_miscent} we are measuring the distance between two quantities with a different definition: in the mock catalogues the center is identified with the BCG, while our estimator depends on the full galaxy distribution on the scale of the cluster. Even if one neglects resolution issues, it is not expected that the two correspond precisely for many of the objects. We note here that it is possible to select BCGs a posteriori from AMICO detections \citep[see][]{2017A&A...598A.107R}.

We also checked for systematics in the amplitude estimate with respect to the centering offset. There is no significant trend as a function of the offset in the mass proxy derived by AMICO. This is expected as, by construction, the algorithm selects the position of the maximum of $A$ as the detection center.

\subsection{Membership}\label{sect:real_members}
\begin{figure}
 \includegraphics[width=\columnwidth]{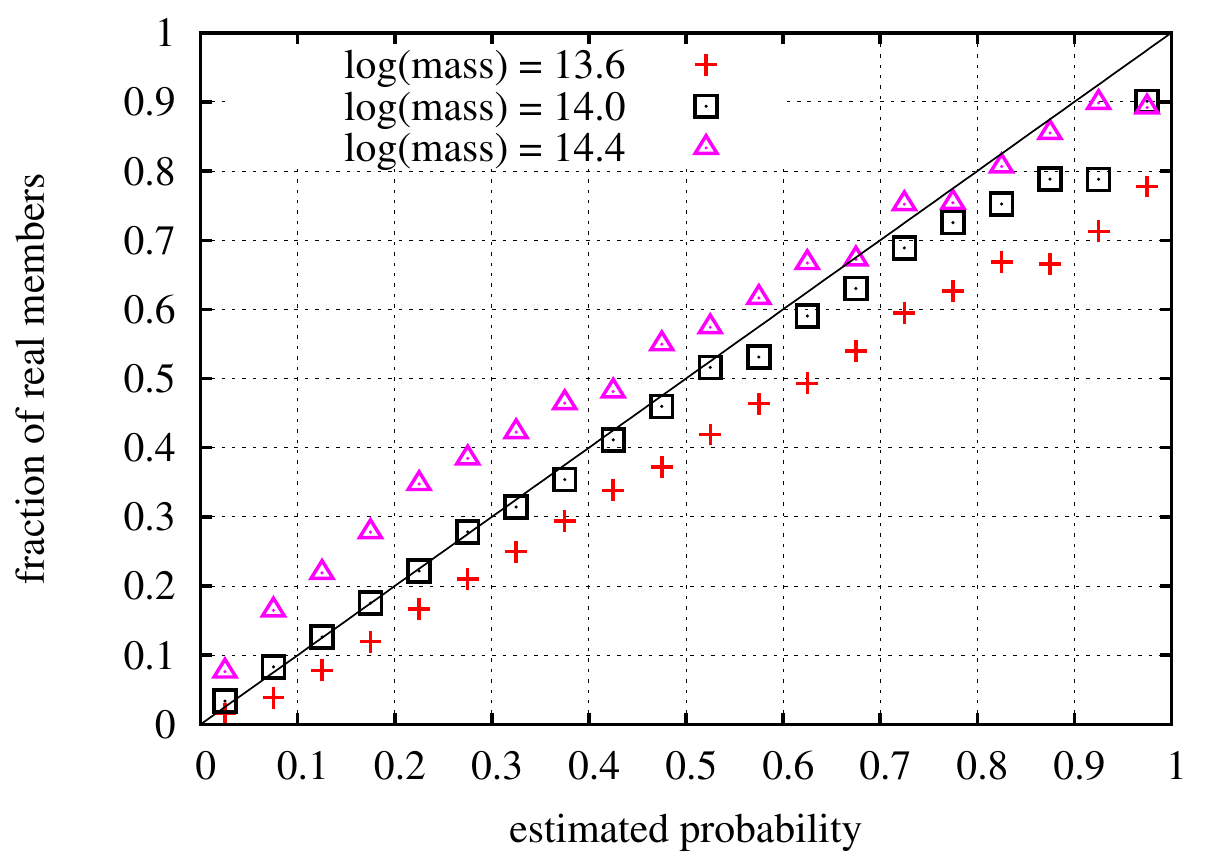}
 \caption{Mean membership probability as estimated by the algorithm versus the fraction of actual members. Red, black and green points refer to galaxy members of haloes with $\log M/(M_\odot/h)$ around 13.6, 14.0 and 14.4, respectively. The error bars are negligible because they are smaller than the point size. The 1-to-1 relation is shown for reference as a black solid line.}
 \label{fig:50_members_mass}
\end{figure}

Having established the performance of the detection procedure, we now focus on the assignment of the membership probability of the galaxies, as we have done in Section \ref{sect:big_members} for the ideal case. Only members of detections matched to real objects were considered in the following analysis. In Figure \ref{fig:50_members_mass} we show the relation between the estimated probability and the fraction of real members for three different bins of mass. As explained in Section \ref{sect:model}, galaxies in the mock catalogues are defined as members when they are at a distance $\leq 1.35 R_{200}$ from the central galaxy. 
The plot in Figure \ref{fig:50_members_mass} shows that, even in a more realistic scenario, AMICO is able to provide a reliable membership assignment through about an order of magnitude in mass. For haloes with masses $\sim 10^{14} M_\odot/h$ the relation is well calibrated with differences lower than 0.1.

The slight over(under)-estimation of the probability for smaller (larger) haloes depends on the mismatch between the model and the actual size of the objects. The flattening of the relation at high probabilities is is mostly due to small miscenterings of the true halo position. To further improve the identification of the members of a cluster, one can a-posteriori correct the membership probability, provided a reasonable estimate of the mass-amplitude relation is obtained, but this goes beyond the scope of this paper.

\subsection{Spurious detections}
In general, the definition of spurious detection has some ambiguity in this case, differently from the simple case analysed in Section \ref{sect:big_false}. We showed in Section \ref{sect:real_massrich}, that, as for any mass proxy based on photometric data, there is a significant scatter between mass and amplitude, and thus also objects of small mass are expected to enter the sample if we do not set a very high threshold in S/N or $A_\text{obs}$. On the other hand, due to the photometric redshift uncertainty, it is necessary to search for detection counterparts in a quite large window in $z$, inside which one will usually find some corresponding halo, if no significant mass threshold is set. This means that in this case there is not an obvious way to discriminate true and false detections, and that most of the information on the quality of the detection method is embedded in the mass-observable plots shown in Figure \ref{fig:50_massrich}.

For this reason, we decided to define the mass threshold for our sample considering the properties the mass-amplitude relation. Given the analytic S/N estimates as a function of amplitude and redshift shown in Figure \ref{fig:analytic_sn}, for each redshift bin we derive the amplitude that corresponds to S/N=2 (see the relation in Fig. 4) and define the value of $M_\text{thr}(z)$ as the one derived from the $A-M$ relation, without considering the scatter. We then consider as valid counterparts only haloes that have mass bigger than $M_\text{thr}(z)$. We note that this mass is actually quite low compared to what is usually defined a cluster and increases as a function of redshift, with $\log M_\text{thr}/(M_\odot/h)$ which goes from 12.44 at $z$ = 0.35 to 13.14 at $z$ = 0.95, following the decrease in S/N for a fixed amplitude. All the detections without a counterpart with $\log M$ > $M_\text{thr}(z)$, are then considered as spurious. 

\begin{figure}
 \includegraphics[width=\columnwidth]{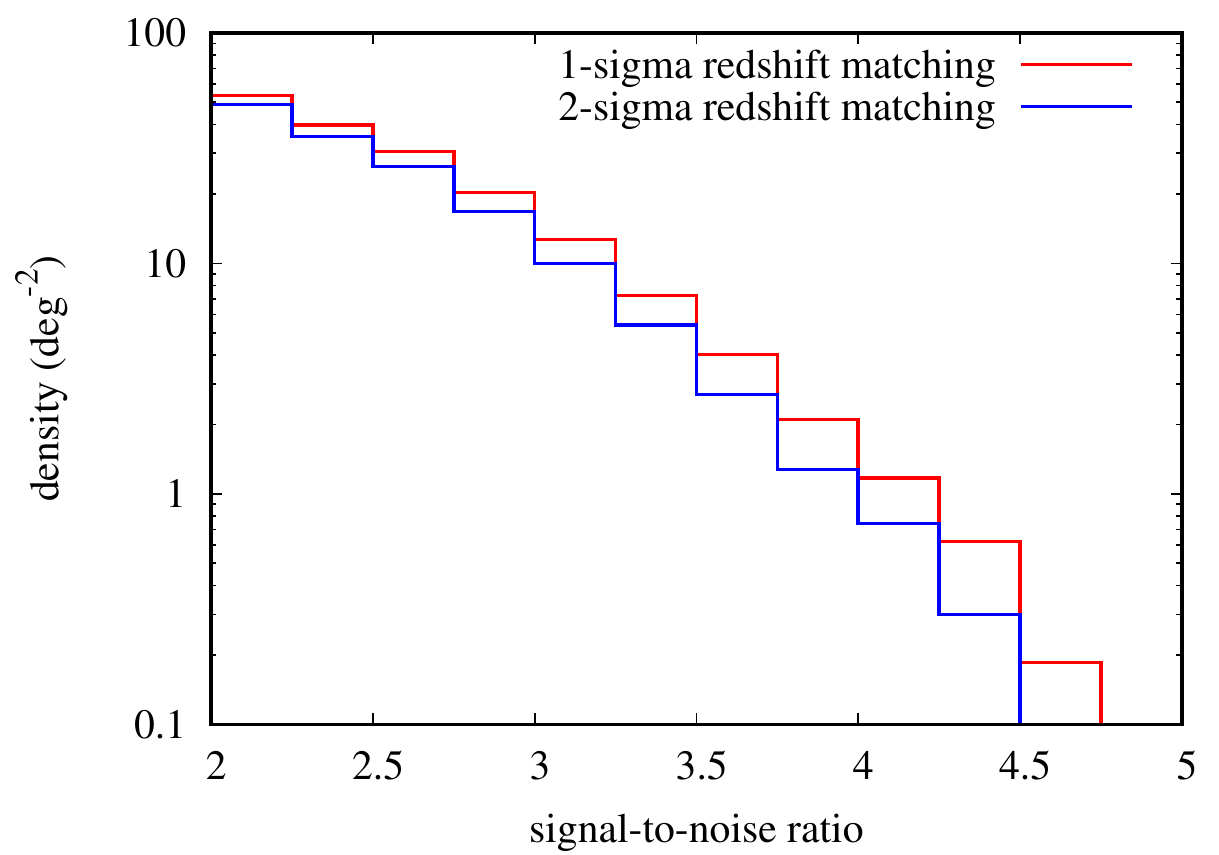}
 \caption{Number of spurious detections per square degree as a function of signal-to-noise ratio. Red line: 1-sigma redshift matching. Blue line: 2-sigma redshift matching.}
 \label{fig:50_false}
\end{figure}

\begin{figure}
 \includegraphics[width=\columnwidth]{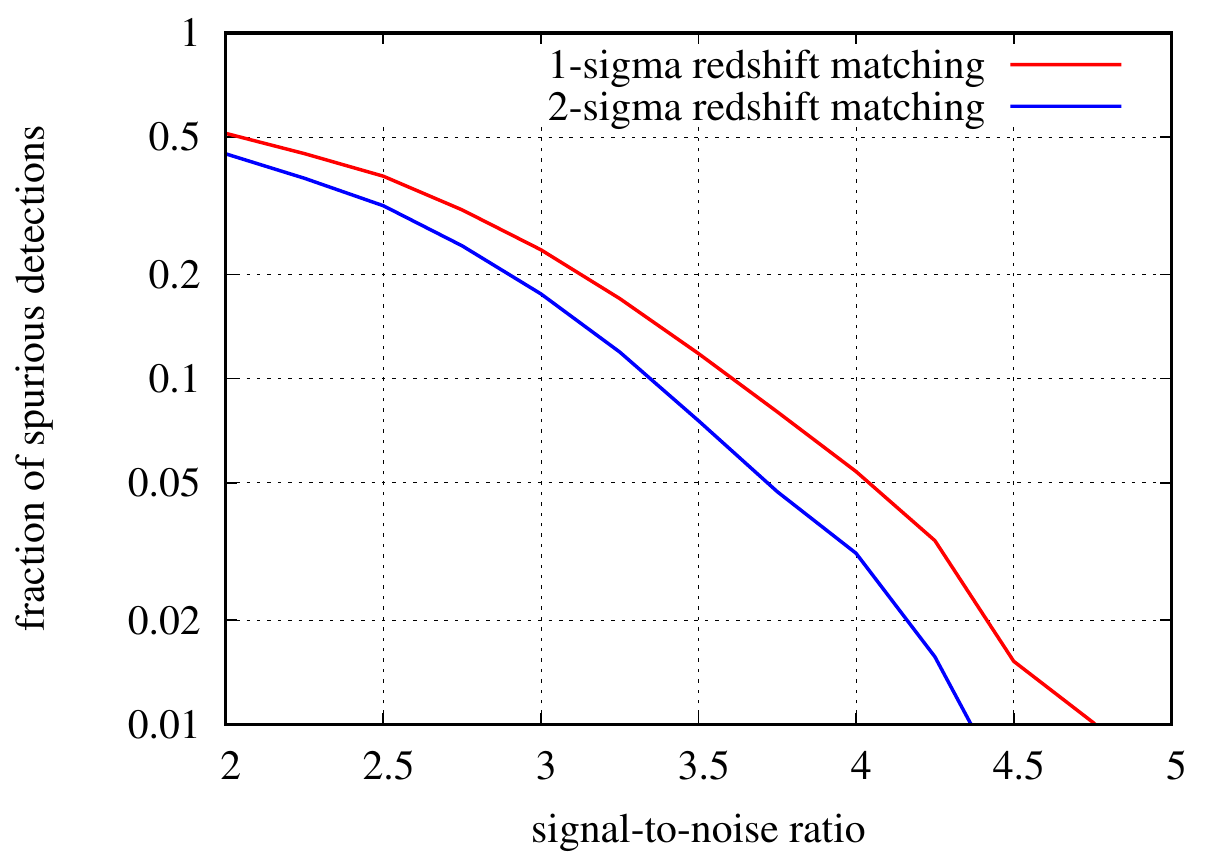}
 \caption{Fraction of spurious detections with a minimum signal-to-noise ratio. Red line: 1-sigma redshift matching. Blue line: 2-sigma redshift matching.}
 \label{fig:50_false_frac}
\end{figure}

We show in Figure \ref{fig:50_false} the density per square degree of spurious detections as a function of their signal-to-noise ratio, and in Figure \ref{fig:50_false_frac} the fraction of unmatched detections with a signal-to-noise larger than a given threshold. Both results are shown for the normal matching procedure (red line) and for the more relaxed criterion of 2-sigma in $z$. The signal-to-noise ratio is confirmed to be a good indicator of the confidence in a detection, as the fraction of spurious detections is monotonically and steeply decreasing. Comparing Figure \ref{fig:50_false} with Figure \ref{fig:big_false}, we see that the number of spurious detections at 2 < S/N < 2.5 is actually lower in this case than in the ideal case of pure random uncorrelated background. This happens because the cleaning procedure applied on high S/N structures removes most of the very low-significance detections. On the other hand, there is a small but not irrelevant number of detections considered as spurious at 3 < S/N < 5. The fraction of unmatched detections goes below $12\%$ ($8 \%$) at S/N > 3.5 for the 1-sigma (2-sigma) redshift matching criterion and becomes negligible at S/N > 5. This value is consistent with what was obtained on COSMOS data in \citet{2011MNRAS.413.1145B} with a previous version of the algorithm which did not implement the deblending procedure and the local correction for local background variations.

We underline again that this definition of spurious detection is somewhat arbitrary. There is no reason why a halo of mass just above $M_\text{thr}$ should be considered as a good counterpart while one just below this threshold should not, given the scatter in the $A-M$ relation. Actually, following the relation between mass and amplitude derived in Section \ref{sect:real_massrich} and its scatter, one can estimate which fraction of objects of a certain mass and redshift is expected to enter the detection catalogue for a given S/N threshold. In our case, we found that this would lead to a large over-estimation of the expected detections, for objects of mass $M \lesssim M_\text{thr}$. In practice, most of the very small objects are eliminated when bigger structures are found and cleaned, because they are physically connected or randomly aligned with them. As a consequence of this, we are actually detecting fewer objects of mass similar or lower than $M_\text{thr}$ with respect to what one would expect combining the mass function and the scatter in the mass-amplitude relation.

\section{Summary and conclusions}\label{sect:concl}
In this work, we presented AMICO (Adaptive Matched Identifier of Clustered Objects), a new implementation of the Optimal Filtering technique for cluster detection presented in \citet{2005A&A...442..851M} and applied to photometric data in \citet{2011MNRAS.413.1145B}. Differently from other approaches, Optimal Filtering is not based on the search of any specific feature of galaxy clusters, but instead aims at maximising the signal-to-noise ratio for the detection of objects which follow a given model in data affected by a given amount of noise. The model for clusters at redshift $z$ can be considered an input parameter in this approach and can in principle be derived from previous observations or iteratively from the data. In this paper, we did not focus on the extraction of the cluster model, and we instead considered the workflow of AMICO once the model is known.

The first step in the detection procedure is the creation of a 3D map constructed through the application of the Optimal Filter to the data. This map contains in each pixel an estimate of the amplitude of a tentative cluster centred in that position in the sky and redshift, together with its S/N and likelihood. The main novelty we introduce with AMICO with respect to \citet{2011MNRAS.413.1145B} is the method to derive the detection catalogue from the filtered map. Instead of just selecting peaks above a certain S/N in the map, we apply a procedure which aims at erasing iteratively the imprint of the identified structures. This helps in the detection of smaller structures partially overlapping with them in the sky. As a by-product, we also obtain for each galaxy a reliable estimate of its probability to be a member of the candidate clusters.

We tested the performance of the algorithm running it blindly on mock catalogues. We first used catalogues with clusters that follow our model and a constant random background to test the code in a controlled environment. In this way we verified that AMICO is able to reach a completeness in agreement with theoretical expectations, and unbiased estimates of amplitude and redshift for the detected objects. All expectations were met. We then turned to specific tests on the capability of AMICO to disentangle structures which are close-by in the sky or aligned along the line of sight. In 50\% of the cases, AMICO proved to be able to distinguish two structures whose relative distance is $\sim 0.5 \times R_{200}$ in the sky or $\sim 2 \times \sigma_z$ in redshift, where $\sigma_z$ is the typical photometric redshift uncertainty of the galaxies.

We finally turned our attention to the analysis of cosmological mocks, produced by semi-analytic modelling of galaxy evolution on a cosmological N-body simulation. In this case the galaxy distribution is more similar to what observed in the Universe, with haloes of different masses embedded in the large-scale structure. This makes the detection procedure more difficult, because the background is not homogeneous and the galaxy distribution of each cluster may have peculiar properties. Despite the complexity of the data, AMICO achieves a completeness for haloes with mass $\sim 10^{14} M_\odot/h$ close to 1 all over the considered redshift range ($0.3 < z < 1.0$), in line with the theoretical predictions and the results on the ideal mocks. The redshift of the detected haloes is recovered with no significant bias, and with a scatter which is of $\sim 0.5 \sigma_z$, higher than in the ideal case.

To analyse in more depth the results of the detection procedure, we performed a fit of the mass-amplitude relation in different redshift bins, considering the mass range for which the sample is (almost) completely detected. The mass-amplitude relation has a (logarithmic) slope between 0.50 and 0.60. Most importantly, the bias for haloes with masses close to the calibration mass of $10^{14} M_\odot/h$ (i.e. the intercept of the $\log A - \log M$ relation) is consistent with zero at all $z$. This shows that the measured amplitude respects the input calibration of the model. Also the logarithmic scatter of the amplitude does not show any evolution with $z$, and is always between 0.13 and 0.15. This indicates that the amplitude $A$ is an unbiased redshift-independent mass proxy, provided that the model is calibrated with haloes of similar mass at each redshift. A key step in this result is the local estimate of the background population, which allows to subtract effectively the contribution of the large-scale structure correlated with the clusters.

Finally, we verified that the S/N computed by AMICO is a reliable indicator of the goodness of a detection: the fraction of false positives is $< 8-12 \%$ at S/N > 3.5 and practically zero at S/N > 5. Note that for this calculation we considered as spurious all the detections which do not have a counterpart in the halo catalogue with mass above $M_\text{thr}(z)$, which corresponds to a theoretical S/N = 2. This sharp truncation in mass is motivated by our approach, but nevertheless it neglects all haloes with $M < M_\text{thr}$ which can be, and actually are, detected because of the scatter in the mass-amplitude relation. Therefore our estimate is conservative.

AMICO is an algorithm that builds on the well-estabilished Optimal Filtering approach for cluster detection, solving in an effective way the problem of extracting detections from the map. The cleaning procedure allows the detection of groups and clusters that are close-by or aligned to larger structures, a particularly challenging task for the peak selection used traditionally in filtering methods \citep{1999ApJ...517...78K,2010MNRAS.406..673M,2011MNRAS.413.1145B}. In some way, the cleaning procedure resembles the so-called \textit{percolation} in redMaPPer \citep{2014ApJ...785..104R}, where a membership probability is attributed to the galaxies in a similar way to our Equations \ref{eq:member_prob} and \ref{eq:field_prob}. Similarly to redMaPPer, the final center, redshift and richness of the detections are computed taking into account higher-ranked close-by clusters. We point out, however, that the way in which detections are performed is different in the two algorithms: redMaPPer computes and ranks the candidate cluster centres on the original catalogue, while in AMICO each cluster candidate is detected only after those with larger likelihood have been erased. Moreover, redMaPPer removes from the list of possible centres the galaxies which have a probability of belonging to an other structure larger than 50\%, while AMICO does not set any arbitrary threshold, as the S/N in each point takes naturally into account the available probability of the surrounding galaxies.

By choice, in this work we ignored the problem of defining a correct cluster model for a given dataset, which is for sure an important part of the challenge of cluster detection, especially when the redshift limits of the surveys are pushed above $z$ = 1, when less information about known structures is available. We are working on the capability of AMICO to extract the cluster model from a given data-set, and we plan to show this feature in a future work. 

Finally we note that, thanks to its flexible structure, AMICO can in principle be applied to other astronomical problems, such as the detection of dwarf galaxies or globular clusters in star catalogues.

\section{Acknowledgements}
FB, MR and LM thank the support from the grants ASI n.I/023/12/0 ``Attivit\`a relative alla fase B2/C per la missione Euclid'' and PRIN MIUR 2015 ``Cosmology and Fundamental Physics: Illuminating the Dark Universe with Euclid''. MM was supported by the SFB-Transregio TR33 ``The Dark Universe''. We thank O. Cucciati and S. Bardelli for the useful discussions.

\bibliographystyle{mnras}
\bibliography{finder}

\begin{thebibliography}{}
\makeatletter
\relax
\def\mn@urlcharsother{\let\do\@makeother \do\$\do\&\do\#\do\^\do\_\do\%\do\~}
\def\mn@doi{\begingroup\mn@urlcharsother \@ifnextchar [ {\mn@doi@}
  {\mn@doi@[]}}
\def\mn@doi@[#1]#2{\def\@tempa{#1}\ifx\@tempa\@empty \href
  {http://dx.doi.org/#2} {doi:#2}\else \href {http://dx.doi.org/#2} {#1}\fi
  \endgroup}
\def\mn@eprint#1#2{\mn@eprint@#1:#2::\@nil}
\def\mn@eprint@arXiv#1{\href {http://arxiv.org/abs/#1} {{\tt arXiv:#1}}}
\def\mn@eprint@dblp#1{\href {http://dblp.uni-trier.de/rec/bibtex/#1.xml}
  {dblp:#1}}
\def\mn@eprint@#1:#2:#3:#4\@nil{\def\@tempa {#1}\def\@tempb {#2}\def\@tempc
  {#3}\ifx \@tempc \@empty \let \@tempc \@tempb \let \@tempb \@tempa \fi \ifx
  \@tempb \@empty \def\@tempb {arXiv}\fi \@ifundefined
  {mn@eprint@\@tempb}{\@tempb:\@tempc}{\expandafter \expandafter \csname
  mn@eprint@\@tempb\endcsname \expandafter{\@tempc}}}

\bibitem[\protect\citeauthoryear{{Abell}}{{Abell}}{1958}]{1958ApJS....3..211A}
{Abell} G.~O.,  1958, \mn@doi [\apjs] {10.1086/190036}, \href
  {http://adsabs.harvard.edu/abs/1958ApJS....3..211A} {3, 211}

\bibitem[\protect\citeauthoryear{{Andreon} \& {Berg{\'e}}}{{Andreon} \&
  {Berg{\'e}}}{2012}]{2012A&A...547A.117A}
{Andreon} S.,  {Berg{\'e}} J.,  2012, \mn@doi [\aap]
  {10.1051/0004-6361/201220115}, \href
  {http://adsabs.harvard.edu/abs/2012A%26A...547A.117A} {547, A117}

\bibitem[\protect\citeauthoryear{{Ascaso}, {Wittman}  \&
  {Ben{\'{\i}}tez}}{{Ascaso} et~al.}{2012}]{2012MNRAS.420.1167A}
{Ascaso} B.,  {Wittman} D.,   {Ben{\'{\i}}tez} N.,  2012, \mn@doi [\mnras]
  {10.1111/j.1365-2966.2011.20107.x}, \href
  {http://adsabs.harvard.edu/abs/2012MNRAS.420.1167A} {420, 1167}

\bibitem[\protect\citeauthoryear{{Bellagamba}, {Maturi}, {Hamana},
  {Meneghetti}, {Miyazaki}  \& {Moscardini}}{{Bellagamba}
  et~al.}{2011}]{2011MNRAS.413.1145B}
{Bellagamba} F.,  {Maturi} M.,  {Hamana} T.,  {Meneghetti} M.,  {Miyazaki} S.,
   {Moscardini} L.,  2011, \mn@doi [\mnras] {10.1111/j.1365-2966.2011.18202.x},
  \href {http://adsabs.harvard.edu/abs/2011MNRAS.413.1145B} {413, 1145}

\bibitem[\protect\citeauthoryear{{Botzler}, {Snigula}, {Bender}  \&
  {Hopp}}{{Botzler} et~al.}{2004}]{2004MNRAS.349..425B}
{Botzler} C.~S.,  {Snigula} J.,  {Bender} R.,   {Hopp} U.,  2004, \mn@doi
  [\mnras] {10.1111/j.1365-2966.2004.07468.x}, \href
  {http://adsabs.harvard.edu/abs/2004MNRAS.349..425B} {349, 425}

\bibitem[\protect\citeauthoryear{{De Lucia} \& {Blaizot}}{{De Lucia} \&
  {Blaizot}}{2007}]{2007MNRAS.375....2D}
{De Lucia} G.,  {Blaizot} J.,  2007, \mn@doi [\mnras]
  {10.1111/j.1365-2966.2006.11287.x}, \href
  {http://adsabs.harvard.edu/abs/2007MNRAS.375....2D} {375, 2}

\bibitem[\protect\citeauthoryear{{Dong}, {Pierpaoli}, {Gunn}  \&
  {Wechsler}}{{Dong} et~al.}{2008}]{2008ApJ...676..868D}
{Dong} F.,  {Pierpaoli} E.,  {Gunn} J.~E.,   {Wechsler} R.~H.,  2008, \mn@doi
  [\apj] {10.1086/522490}, \href
  {http://adsabs.harvard.edu/abs/2008ApJ...676..868D} {676, 868}

\bibitem[\protect\citeauthoryear{{Gladders} \& {Yee}}{{Gladders} \&
  {Yee}}{2000}]{2000AJ....120.2148G}
{Gladders} M.~D.,  {Yee} H.~K.~C.,  2000, \mn@doi [\aj] {10.1086/301557}, \href
  {http://adsabs.harvard.edu/abs/2000AJ....120.2148G} {120, 2148}

\bibitem[\protect\citeauthoryear{{Hansen}, {McKay}, {Wechsler}, {Annis},
  {Sheldon}  \& {Kimball}}{{Hansen} et~al.}{2005}]{2005ApJ...633..122H}
{Hansen} S.~M.,  {McKay} T.~A.,  {Wechsler} R.~H.,  {Annis} J.,  {Sheldon}
  E.~S.,   {Kimball} A.,  2005, \mn@doi [\apj] {10.1086/444554}, \href
  {http://adsabs.harvard.edu/abs/2005ApJ...633..122H} {633, 122}

\bibitem[\protect\citeauthoryear{{Kaiser} et~al.,}{{Kaiser}
  et~al.}{2002}]{2002SPIE.4836..154K}
{Kaiser} N.,  et~al., 2002, in {Tyson} J.~A.,  {Wolff} S.,  eds,  \procspie
  Vol. 4836, Survey and Other Telescope Technologies and Discoveries. pp
  154--164, \mn@doi{10.1117/12.457365}

\bibitem[\protect\citeauthoryear{{Kepner}, {Fan}, {Bahcall}, {Gunn}, {Lupton}
  \& {Xu}}{{Kepner} et~al.}{1999}]{1999ApJ...517...78K}
{Kepner} J.,  {Fan} X.,  {Bahcall} N.,  {Gunn} J.,  {Lupton} R.,   {Xu} G.,
  1999, \mn@doi [\apj] {10.1086/307160}, \href
  {http://adsabs.harvard.edu/abs/1999ApJ...517...78K} {517, 78}

\bibitem[\protect\citeauthoryear{{Koester} et~al.,}{{Koester}
  et~al.}{2007}]{2007ApJ...660..221K}
{Koester} B.~P.,  et~al., 2007, \mn@doi [\apj] {10.1086/512092}, \href
  {http://adsabs.harvard.edu/abs/2007ApJ...660..221K} {660, 221}

\bibitem[\protect\citeauthoryear{{LSST Science Collaboration} et~al.,}{{LSST
  Science Collaboration} et~al.}{2009}]{2009arXiv0912.0201L}
{LSST Science Collaboration} et~al., 2009, preprint, \href
  {http://adsabs.harvard.edu/abs/2009arXiv0912.0201L} {} (\mn@eprint {arXiv}
  {0912.0201})

\bibitem[\protect\citeauthoryear{{Laureijs} et~al.,}{{Laureijs}
  et~al.}{2011}]{2011arXiv1110.3193L}
{Laureijs} R.,  et~al., 2011, preprint, \href
  {http://adsabs.harvard.edu/abs/2011arXiv1110.3193L} {} (\mn@eprint {arXiv}
  {1110.3193})

\bibitem[\protect\citeauthoryear{{Le Fevre} et~al.,}{{Le Fevre}
  et~al.}{2003}]{2003SPIE.4834..173L}
{Le Fevre} O.,  et~al., 2003, in {Guhathakurta} P.,  ed.,  \procspie Vol. 4834,
  Discoveries and Research Prospects from 6- to 10-Meter-Class Telescopes II.
  pp 173--182, \mn@doi{10.1117/12.457547}

\bibitem[\protect\citeauthoryear{{Licitra}, {Mei}, {Raichoor}, {Erben}  \&
  {Hildebrandt}}{{Licitra} et~al.}{2016}]{2016MNRAS.455.3020L}
{Licitra} R.,  {Mei} S.,  {Raichoor} A.,  {Erben} T.,   {Hildebrandt} H.,
  2016, \mn@doi [\mnras] {10.1093/mnras/stv2309}, \href
  {http://adsabs.harvard.edu/abs/2016MNRAS.455.3020L} {455, 3020}

\bibitem[\protect\citeauthoryear{{Lumsden}, {Nichol}, {Collins}  \&
  {Guzzo}}{{Lumsden} et~al.}{1992}]{1992MNRAS.258....1L}
{Lumsden} S.~L.,  {Nichol} R.~C.,  {Collins} C.~A.,   {Guzzo} L.,  1992,
  \mn@doi [\mnras] {10.1093/mnras/258.1.1}, \href
  {http://adsabs.harvard.edu/abs/1992MNRAS.258....1L} {258, 1}

\bibitem[\protect\citeauthoryear{{Maturi}, {Meneghetti}, {Bartelmann}, {Dolag}
  \& {Moscardini}}{{Maturi} et~al.}{2005}]{2005A&A...442..851M}
{Maturi} M.,  {Meneghetti} M.,  {Bartelmann} M.,  {Dolag} K.,   {Moscardini}
  L.,  2005, \mn@doi [\aap] {10.1051/0004-6361:20042600}, \href
  {http://adsabs.harvard.edu/abs/2005A%26A...442..851M} {442, 851}

\bibitem[\protect\citeauthoryear{{Meneux} et~al.,}{{Meneux}
  et~al.}{2008}]{2008A&A...478..299M}
{Meneux} B.,  et~al., 2008, \mn@doi [\aap] {10.1051/0004-6361:20078182}, \href
  {http://adsabs.harvard.edu/abs/2008A%26A...478..299M} {478, 299}

\bibitem[\protect\citeauthoryear{{Milkeraitis}, {van Waerbeke}, {Heymans},
  {Hildebrandt}, {Dietrich}  \& {Erben}}{{Milkeraitis}
  et~al.}{2010}]{2010MNRAS.406..673M}
{Milkeraitis} M.,  {van Waerbeke} L.,  {Heymans} C.,  {Hildebrandt} H.,
  {Dietrich} J.~P.,   {Erben} T.,  2010, \mn@doi [\mnras]
  {10.1111/j.1365-2966.2010.16720.x}, \href
  {http://adsabs.harvard.edu/abs/2010MNRAS.406..673M} {406, 673}

\bibitem[\protect\citeauthoryear{{Navarro}, {Frenk}  \& {White}}{{Navarro}
  et~al.}{1997}]{1997ApJ...490..493N}
{Navarro} J.~F.,  {Frenk} C.~S.,   {White} S.~D.~M.,  1997, \apj, \href
  {http://adsabs.harvard.edu/abs/1997ApJ...490..493N} {490, 493}

\bibitem[\protect\citeauthoryear{{Pace}, {Maturi}, {Meneghetti}, {Bartelmann},
  {Moscardini}  \& {Dolag}}{{Pace} et~al.}{2007}]{2007A&A...471..731P}
{Pace} F.,  {Maturi} M.,  {Meneghetti} M.,  {Bartelmann} M.,  {Moscardini} L.,
   {Dolag} K.,  2007, \mn@doi [\aap] {10.1051/0004-6361:20077217}, \href
  {http://adsabs.harvard.edu/abs/2007A%26A...471..731P} {471, 731}

\bibitem[\protect\citeauthoryear{{Pace}, {Maturi}, {Bartelmann}, {Cappelluti},
  {Dolag}, {Meneghetti}  \& {Moscardini}}{{Pace}
  et~al.}{2008}]{2008A&A...483..389P}
{Pace} F.,  {Maturi} M.,  {Bartelmann} M.,  {Cappelluti} N.,  {Dolag} K.,
  {Meneghetti} M.,   {Moscardini} L.,  2008, \mn@doi [\aap]
  {10.1051/0004-6361:200809550}, \href
  {http://adsabs.harvard.edu/abs/2008A%26A...483..389P} {483, 389}

\bibitem[\protect\citeauthoryear{{Postman}, {Lubin}, {Gunn}, {Oke}, {Hoessel},
  {Schneider}  \& {Christensen}}{{Postman} et~al.}{1996}]{1996AJ....111..615P}
{Postman} M.,  {Lubin} L.~M.,  {Gunn} J.~E.,  {Oke} J.~B.,  {Hoessel} J.~G.,
  {Schneider} D.~P.,   {Christensen} J.~A.,  1996, \mn@doi [\aj]
  {10.1086/117811}, \href {http://adsabs.harvard.edu/abs/1996AJ....111..615P}
  {111, 615}

\bibitem[\protect\citeauthoryear{{Radovich} et~al.,}{{Radovich}
  et~al.}{2017}]{2017A&A...598A.107R}
{Radovich} M.,  et~al., 2017, \mn@doi [\aap] {10.1051/0004-6361/201629353},
  \href {http://adsabs.harvard.edu/abs/2017A%26A...598A.107R} {598, A107}

\bibitem[\protect\citeauthoryear{{Ramella}, {Nonino}, {Boschin}  \&
  {Fadda}}{{Ramella} et~al.}{1999}]{1999ASPC..176..108R}
{Ramella} M.,  {Nonino} M.,  {Boschin} W.,   {Fadda} D.,  1999, in {Giuricin}
  G.,  {Mezzetti} M.,   {Salucci} P.,  eds,  Astronomical Society of the
  Pacific Conference Series Vol. 176, Observational Cosmology: The Development
  of Galaxy Systems. p.~108 (\mn@eprint {} {astro-ph/9810124})

\bibitem[\protect\citeauthoryear{{Rykoff} et~al.,}{{Rykoff}
  et~al.}{2012}]{2012ApJ...746..178R}
{Rykoff} E.~S.,  et~al., 2012, \mn@doi [\apj] {10.1088/0004-637X/746/2/178},
  \href {http://adsabs.harvard.edu/abs/2012ApJ...746..178R} {746, 178}

\bibitem[\protect\citeauthoryear{{Rykoff} et~al.,}{{Rykoff}
  et~al.}{2014}]{2014ApJ...785..104R}
{Rykoff} E.~S.,  et~al., 2014, \mn@doi [\apj] {10.1088/0004-637X/785/2/104},
  \href {http://adsabs.harvard.edu/abs/2014ApJ...785..104R} {785, 104}

\bibitem[\protect\citeauthoryear{{Rykoff} et~al.,}{{Rykoff}
  et~al.}{2016}]{2016ApJS..224....1R}
{Rykoff} E.~S.,  et~al., 2016, \mn@doi [\apjs] {10.3847/0067-0049/224/1/1},
  \href {http://adsabs.harvard.edu/abs/2016ApJS..224....1R} {224, 1}

\bibitem[\protect\citeauthoryear{{Schechter}}{{Schechter}}{1976}]{1976ApJ...203..297S}
{Schechter} P.,  1976, \mn@doi [\apj] {10.1086/154079}, \href
  {http://adsabs.harvard.edu/abs/1976ApJ...203..297S} {203, 297}

\bibitem[\protect\citeauthoryear{{Sereno} \& {Ettori}}{{Sereno} \&
  {Ettori}}{2015}]{2015MNRAS.450.3633S}
{Sereno} M.,  {Ettori} S.,  2015, \mn@doi [\mnras] {10.1093/mnras/stv810},
  \href {http://adsabs.harvard.edu/abs/2015MNRAS.450.3633S} {450, 3633}

\bibitem[\protect\citeauthoryear{{Sereno} \& {Ettori}}{{Sereno} \&
  {Ettori}}{2017}]{2017MNRAS.468.3322S}
{Sereno} M.,  {Ettori} S.,  2017, \mn@doi [\mnras] {10.1093/mnras/stx576},
  \href {http://adsabs.harvard.edu/abs/2017MNRAS.468.3322S} {468, 3322}

\bibitem[\protect\citeauthoryear{{Shapley}}{{Shapley}}{1933}]{1933PNAS...19..591S}
{Shapley} H.,  1933, \mn@doi [Proceedings of the National Academy of Science]
  {10.1073/pnas.19.6.591}, \href
  {http://adsabs.harvard.edu/abs/1933PNAS...19..591S} {19, 591}

\bibitem[\protect\citeauthoryear{{Shectman}}{{Shectman}}{1985}]{1985ApJS...57...77S}
{Shectman} S.~A.,  1985, \mn@doi [\apjs] {10.1086/190996}, \href
  {http://adsabs.harvard.edu/abs/1985ApJS...57...77S} {57, 77}

\bibitem[\protect\citeauthoryear{{Simet}, {McClintock}, {Mandelbaum}, {Rozo},
  {Rykoff}, {Sheldon}  \& {Wechsler}}{{Simet}
  et~al.}{2017}]{2017MNRAS.466.3103S}
{Simet} M.,  {McClintock} T.,  {Mandelbaum} R.,  {Rozo} E.,  {Rykoff} E.,
  {Sheldon} E.,   {Wechsler} R.~H.,  2017, \mn@doi [\mnras]
  {10.1093/mnras/stw3250}, \href
  {http://adsabs.harvard.edu/abs/2017MNRAS.466.3103S} {466, 3103}

\bibitem[\protect\citeauthoryear{{Soares-Santos} et~al.,}{{Soares-Santos}
  et~al.}{2011}]{2011ApJ...727...45S}
{Soares-Santos} M.,  et~al., 2011, \mn@doi [\apj] {10.1088/0004-637X/727/1/45},
  \href {http://adsabs.harvard.edu/abs/2011ApJ...727...45S} {727, 45}

\bibitem[\protect\citeauthoryear{{Springel} et~al.,}{{Springel}
  et~al.}{2005}]{2005Natur.435..629S}
{Springel} V.,  et~al., 2005, \mn@doi [\nat] {10.1038/nature03597}, \href
  {http://adsabs.harvard.edu/abs/2005Natur.435..629S} {435, 629}

\bibitem[\protect\citeauthoryear{{The Dark Energy Survey Collaboration}}{{The
  Dark Energy Survey Collaboration}}{2005}]{2005astro.ph.10346T}
{The Dark Energy Survey Collaboration} 2005, preprint, \href
  {http://adsabs.harvard.edu/abs/2005astro.ph.10346T} {} (\mn@eprint {}
  {0510346})

\bibitem[\protect\citeauthoryear{{Trevese}, {Castellano}, {Fontana}  \&
  {Giallongo}}{{Trevese} et~al.}{2007}]{2007A&A...463..853T}
{Trevese} D.,  {Castellano} M.,  {Fontana} A.,   {Giallongo} E.,  2007, \mn@doi
  [\aap] {10.1051/0004-6361:20065384}, \href
  {http://adsabs.harvard.edu/abs/2007A%26A...463..853T} {463, 853}

\bibitem[\protect\citeauthoryear{{Viola}, {Maturi}  \& {Bartelmann}}{{Viola}
  et~al.}{2010}]{2010MNRAS.403..859V}
{Viola} M.,  {Maturi} M.,   {Bartelmann} M.,  2010, \mn@doi [\mnras]
  {10.1111/j.1365-2966.2009.16165.x}, \href
  {http://adsabs.harvard.edu/abs/2010MNRAS.403..859V} {403, 859}

\bibitem[\protect\citeauthoryear{{Wen}, {Han}  \& {Liu}}{{Wen}
  et~al.}{2012}]{2012ApJS..199...34W}
{Wen} Z.~L.,  {Han} J.~L.,   {Liu} F.~S.,  2012, \mn@doi [\apjs]
  {10.1088/0067-0049/199/2/34}, \href
  {http://adsabs.harvard.edu/abs/2012ApJS..199...34W} {199, 34}

\bibitem[\protect\citeauthoryear{{White} \& {Kochanek}}{{White} \&
  {Kochanek}}{2002}]{2002ApJ...574...24W}
{White} M.,  {Kochanek} C.~S.,  2002, \mn@doi [\apj] {10.1086/340944}, \href
  {http://adsabs.harvard.edu/abs/2002ApJ...574...24W} {574, 24}

\bibitem[\protect\citeauthoryear{{de Jong} et~al.,}{{de Jong}
  et~al.}{2013}]{2013Msngr.154...44J}
{de Jong} J.~T.~A.,  et~al., 2013, The Messenger, \href
  {http://adsabs.harvard.edu/abs/2013Msngr.154...44J} {154, 44}

\bibitem[\protect\citeauthoryear{{van Uitert}, {Gilbank}, {Hoekstra},
  {Semboloni}, {Gladders}  \& {Yee}}{{van Uitert}
  et~al.}{2016}]{2016A&A...586A..43V}
{van Uitert} E.,  {Gilbank} D.~G.,  {Hoekstra} H.,  {Semboloni} E.,  {Gladders}
  M.~D.,   {Yee} H.~K.~C.,  2016, \mn@doi [\aap] {10.1051/0004-6361/201526719},
  \href {http://adsabs.harvard.edu/abs/2016A%26A...586A..43V} {586, A43}

\makeatother
\end{thebibliography}

\bsp	
\label{lastpage}
\end{document}